\documentclass[%
 aip,
 jmp,%
 amsmath,amssymb,
reprint,%
]{revtex4-1}

\usepackage[colorlinks,linkcolor=blue,citecolor=blue,urlcolor=black,bookmarks=false,hypertexnames=true]{hyperref} 
\usepackage{amsmath}
\usepackage{xspace}
\usepackage{multirow}
\usepackage{latexsym}
\usepackage{braket}
\usepackage{cleveref}
\usepackage{enumerate}
\usepackage{mathrsfs}
\usepackage{enumitem}
\usepackage{color}
\usepackage[version=3]{mhchem}
\usepackage{caption,setspace}
\usepackage{graphicx}
\usepackage{subfigure}
\usepackage{threeparttable}
\usepackage{url}

\allowdisplaybreaks
\usepackage{array}
\newcolumntype{L}[1]{>{\raggedright\let\newline\\\arraybackslash\hspace{0pt}}m{#1}}
\newcolumntype{C}[1]{>{\centering\let\newline\\\arraybackslash\hspace{0pt}}m{#1}}
\newcolumntype{R}[1]{>{\raggedleft\let\newline\\\arraybackslash\hspace{0pt}}m{#1}}

\renewcommand{\c}[1]{a^\dagger_{#1}}
\renewcommand{\a}[1]{a_{#1}}

\newcommand{\om}{\omega}

\newcommand{\SC}{\ensuremath{\Delta S^{2}}\xspace}

\crefname{figure}{Figure}{Figures}
\crefname{table}{Table}{Tables}
\crefname{equation}{Eq.}{Eqs.}
\crefname{section}{Section}{Sections}

\newcommand{\cm}{\ensuremath{\text{cm}^{-1}}\xspace}

\renewcommand{\a}[1]{a^{\ }_{#1}}

\newcommand*{\mae}{$\mathrm{MAE}$\xspace}
\newcommand*{\std}{$\mathrm{STDV}$\xspace}

\begin{document}

\author{Terrence~L.~Stahl}
\affiliation{Department of Chemistry and Biochemistry, The Ohio State University, Columbus, Ohio 43210, USA}

\author{Alexander~Yu.~Sokolov}
\email{sokolov.8@osu.edu}
\affiliation{Department of Chemistry and Biochemistry, The Ohio State University, Columbus, Ohio 43210, USA}

\title{
%%%%%ArXiv%%%%%
\color{blue}
Quantifying spin contamination in algebraic diagrammatic construction theory of electronic excitations
\vspace{0.25cm}
%%%%%ArXiv%%%%%
}

\begin{abstract}
Algebraic diagrammatic construction (ADC) is a computationally efficient approach for simulating excited electronic states, absorption spectra, and electron correlation.
Due to their origin in perturbation theory, the single-reference ADC methods may be susceptible to spin contamination when applied to molecules with unpaired electrons.
In this work, we develop an approach to quantify spin contamination in the ADC calculations of electronic excitations and apply it to a variety of open-shell molecules starting with either the unrestricted (UHF) or restricted open-shell (ROHF) Hartree--Fock reference wavefunctions. 
Our results show that the accuracy of low-order ADC approximations (ADC(2), ADC(3))  significantly decreases when the UHF reference spin contamination exceeds 0.05 a.u.
Such strongly spin-contaminated molecules exhibit severe excited-state spin symmetry breaking that contributes to decreasing the quality of computed excitation energies and oscillator strengths. 
In a case study of phenyl radical, we demonstrate that spin contamination can significantly affect the simulated UV/Vis spectra, altering the relative energies, intensities, and order of electronic transitions.
The results presented here motivate the development of spin-adapted ADC methods for open-shell molecules.
\end{abstract}

\titlepage

\maketitle

%%%%%%%%%%%%%%%%%%%%%%%%%%%%
% Introduction              
%%%%%%%%%%%%%%%%%%%%%%%%%%%%
\section{Introduction}
\label{sec:intro}
Molecules with unpaired electrons are the key intermediates in a variety of chemical processes that are important in environmental and combustion chemistry, organic synthesis, catalysis, and biochemistry.\cite{tyndall2001atmospheric,martin2007electronic,chhantyal2010observation,
liou2010reactive,poljsak2011skin,narendrapurapu2011combustion,gligorovski2015environmental,
michelsen2017probing,li2018tddft,li2019computational,wang2019wide,wilcox2018stable,franke2019tert,frances2020photodissociation,wang2022overlooked} 
A powerful tool to identify the presence of these short-lived chemical species is the UV/Vis absorption spectroscopy, which combines high sensitivity and specificity, allowing to detect transient species in low concentrations. 
However, compared to stable closed-shell molecules, the UV/Vis spectra of molecular radicals are much less understood and can be difficult to interpret without the insight from accurate theoretical calculations. 

Meanwhile, simulating UV/Vis spectra of open-shell molecules is often challenging as they tend to exhibit complicated ground- or excited-state electronic structures, which require rigorous treatments of electron correlation, spin, and orbital relaxation.
Open-shell electronic states can be accurately described using multireference approaches\cite{Buenker:1974p33,Siegbahn:1980p1647,Knowles:1985p259,Wolinski:1987p225,Mukherjee:1977p955,Jeziorski:1981p1668,Werner:1988p5803,Mahapatra:1998p157,Pittner:2003p10876,Evangelista:2007p024102,Datta:2011p214116,Datta:2012p204107,Evangelista:2011p114102,Kohn:2012p176,Nooijen:2014p081102,Samanta:2014p134108,Aoto:2016p074103} that correctly treat their spin and spatial symmetry by incorporating many electronic configurations in the reference wavefunction. 
Alternatively, non-perturbative single-reference methods, such as configuration interaction\cite{sherrill:1999p143,sharma:2017p1595,kallay:2004p9257,kossoski:2023p2258,maurice:1995p361,head:1995p114,coe:2016p1772,maurice:1996p6131,roemelt:2013p3069,handy:1979p426} 
or coupled cluster theory,\cite{Crawford:2000p33,Shavitt:2009,Geertsen:1989p57,Comeau:1993p414,Stanton:1993p7029,Krylov:2008p433,Sekino:1984p255,Koch:1990p3345,Koch:1990p3333,Sattelmeyer:2001p499,Christiansen:2005p106,Sneskov:2012p566,Eriksen:2014p174114} 
have been shown to provide accurate results for open-shell systems when incorporating high-order correlation effects (e.g., triple or quadruple excitations) or utilizing orbitally relaxed non-canonical reference wavefunctions.\cite{scuseria:1987p354,krylov:1998p10669,Sherrill:1998p4171,lochan:2007p164101,bozkaya:2012p204114,Bozkaya:2013p54104,copan2018linear,kossoski2021excited,Kohn:2005p084116} 
However, the computational cost of these methodologies scales steeply with the number of electrons and orbitals in a molecule, significantly constraining their applications.
Lower-cost single-reference approaches for simulating UV/Vis spectra of open-shell molecules have been also developed,\cite{Sneskov:2012p566,Christiansen:1995p409,Hattig:2000p5154,Hattig:2005p37,Helmich:2013p084114,Helmich:2014p35,Stanton:1995p1064,Gwaltney:1996p189,Head-Gordon:1999p593,Schirmer:1982p2395,Schirmer:1991p4647,trofimov1995efficient,Mertins:1996p2140,trofimov1999consistent,Schirmer:2004p11449,Dreuw:2014p82,Sokolov:2018p204113} but their accuracy strongly depends on the quality of Hartree--Fock (HF) reference wavefunction that is employed as the zeroth-order approximation.

Of particular concern is the unphysical violation of spin symmetry, known as spin contamination,\cite{montoya2000spin,szalay2000spin,krylov:2017p151,menon2008consequences,ipatov2009excited,zhang2015spin,kitsaras:2021p131101}  which is frequently encountered in the unrestricted HF (UHF) calculations  of open-shell ground-state wavefunctions and is even more common for excited states.
Spin contamination can give rise to large errors in computed excitation energies, unphysical transition intensities, and incorrect description of state crossings, leading to inaccurate UV/Vis spectra.
Although spin contamination is inherent in open-shell calculations using most single-reference methods, it is particularly detrimental for post-HF approaches based on low-order perturbation theory. 

In this work, we investigate spin contamination in the calculations of excited states and UV/Vis spectra of open-shell molecules using algebraic diagrammatic construction theory (ADC).\cite{Schirmer:1982p2395,Schirmer:1991p4647,trofimov1995efficient,Mertins:1996p2140,trofimov1999consistent,Schirmer:2004p11449,Dreuw:2014p82,Sokolov:2018p204113} 
ADC obtains excitation energies and transition intensities by perturbatively approximating the poles and residues of the polarization propagator and is one of the most widely used theoretical approaches for calculating excited states.
In particular, the second-order ADC method (ADC(2)) has been extensively used for calculating the UV/Vis spectra of large molecules\cite{starcke2009unrestricted,lunkenheimer:2013p977,harbach2014third,sarkar:2020p1567,wong:2021p6703,winter:2013p6623,li2014comparison,panda:2013p2181,plasser:2012p2777,aquino:2011p1217,hofener:2019p3160}
due to its relatively low computational cost and ability to incorporate electron correlation effects.
More accurate and expensive ADC(3)\cite{trofimov1999consistent,trofimov2002electron,harbach2014third,Dreuw:2014p82,maier2023consistent} and ADC(4)\cite{leitner:2022p157} approximations have been also developed.

Although the performance of low-order ADC methods for open-shell molecules has been studied,\cite{starcke2009unrestricted,lefrancois2015adapting,paran2022spin} it is yet unclear how significant is spin contamination (SC) in its calculations when starting from the unrestricted (UHF) or restricted open-shell HF (ROHF) reference wavefunctions. 
Does the extent of SC in UHF reference wavefunction correlate with the magnitude of SC in excited states computed using ADC?
Is it possible to reduce the excited-state SC in ADC calculations by starting with the ROHF reference wavefunction?
Does SC reduce with increasing order of ADC approximation?
And, does it affect the accuracy of ADC oscillator strengths and UV/Vis spectra?
Here, we aim to answer these questions by implementing the calculation of spin operator expectation values in ADC methods and quantifying the SC in neutral excited states of small open-shell molecules. 
This work is a continuation of our earlier study where we investigated the SC in ADC calculations of charged excitations.\cite{stahl2022quantifying} 
We also note that the ADC methods are closely related to the CC$n$ approximations,\cite{Sneskov:2012p566,Christiansen:1995p409,Hattig:2000p5154,Hattig:2005p37} for which SC has been recently studied in Ref.\@ \citenum{kitsaras:2021p131101}.

This paper is organized as follows. 
First, we review ADC for the polarization propagator and describe an approach for calculating spin expectation values and contamination (\cref{sec:theory}).
Next, following the description of our implementation and computational details (\cref{sec:Imp/comp_details}), we present and discuss our calculations of SC in small open-shell molecules, correlating the errors in spin expectation values with the errors in computed excitation energies (\cref{sec:results:energies}) and oscillator strengths (\cref{sec:results:osc_strengths}). 
Finally, in \cref{sec:results:phenyl}, we carry out a study of phenyl radical UV/Vis spectrum where we analyze the effect of SC on the energies and intensities of individual electronic transitions. 

\section{Theory}
\label{sec:theory}

\subsection{Algebraic diagrammatic construction for the polarization propagator}
\label{sec:theory:Propagator_ADC_Theory}

The electronic excitations in linear UV/Vis spectroscopy are characterized by the polarization propagator\cite{csanak1971green,Fetter2003,Dickhoff2008,danovich:2011p377} 
\begin{align}
\label{eqn:polar}
&\Pi_{pq,rs}(\omega)=\sum_{n\neq0}\frac{\braket{\Psi_{0}^{N} |a_{q}^{\dagger}a_{p}| \Psi_{n}^{N}}\braket{\Psi_{n}^{N}|a_{r}^{\dagger}a_{s}| \Psi_{0}^{N}}}{\omega+E_{0}^{N}-E_{n}^{N}} \notag \\ 
&+ \sum_{n\neq0}\frac{\braket{\Psi_{0}^{N} |a_{r}^{\dagger}a_{s}| \Psi_{n}^{N}}\braket{\Psi_{n}^{N}|a_{q}^{\dagger}a_{p}| \Psi_{0}^{N}}}{-\omega+E_{0}^{N}-E_{n}^{N}}\notag \\
&=\Pi_{pq,rs}^{+}(\omega) + \Pi_{qp,sr}^{-}(\omega)
\end{align}
that describes how the $N$-electron chemical system in state $\ket{\Psi_{0}^{N}}$ with energy $E_{0}^{N}$ (usually, the ground state) interacts with a periodic electric field with frequency $\omega$.
In \cref{eqn:polar}, the two components of $\Pi_{pq,rs}(\omega)$, known as the forward ($\Pi_{pq,rs}^{+}(\omega)$) and backward ($\Pi_{qp,sr}^{-}(\omega)$) polarization propagators, are expressed in terms of the exact wavefunctions ($\ket{\Psi_{n}^{N}}$) and energies ($E_{n}^{N}$) of all electronic states. 
In this so-called spectral (or Lehmann) representation, each contribution to $\Pi_{pq,rs}(\omega)$ can be written more compactly as
\begin{align}
	\label{eqn:poladiag}
	\bold{\Pi}^{\pm}(\omega) = \tilde{\bold{X}}^{\pm}(\omega\bold{1}-\tilde{\bold{\Omega}}^{\pm})^{-1} \tilde{\bold{X}}^{\pm\dagger} 
\end{align}
where $\tilde{\bold{X}}^{\pm}$ are the matrices of spectroscopic amplitudes ($\tilde{X}^{+}_{pq,n}=\braket{\Psi_{0}^{N} |a_{q}^{\dagger} a_{p}| \Psi_{n}^{N}}$, $\tilde{X}^{-}_{rs,n}=\braket{\Psi_{0}^{N} |a_{r}^{\dagger}a_{s}| \Psi_{n}^{N}}$) describing the probability of electronic transitions between two molecular orbitals and $\bold{\tilde{\Omega}}^{\pm}$ is the diagonal matrix of transition energies 
with elements $\tilde{\Omega}^{+}_{n} = E_{n}^{N}-E_{0}^{N}$ or $\tilde{\Omega}^{-}_{n} = -E_{n}^{N} + E_{0}^{N}$.

In algebraic diagrammatic construction (ADC),\cite{Schirmer:1982p2395,Schirmer:1991p4647,trofimov1995efficient,Mertins:1996p2140,trofimov1999consistent,Schirmer:2004p11449,danovich:2011p377,Dreuw:2014p82,Sokolov:2018p204113} the excitation energies and transition probabilities in UV/Vis spectra are computed by approximating $\tilde{\bold{X}}^{\pm}$  and $\bold{\tilde{\Omega}}^{\pm}$  using perturbation theory.
These approximations are derived by ensuring that at any perturbation order $\bold{\Pi}^{+}(\omega)$ and $\bold{\Pi}^{-}(\omega)$ are fully decoupled from each other and can be treated independently.
Due to the time-reversal symmetry, the backward polarization propagator can be obtained as $\bold{\Pi}^{-}(\omega) = (\bold{\Pi}^{+}(-\omega))^\dag$, and does not need to be considered explicitly.
For this reason, ADC focuses on the forward component, expressing it in the non-diagonal matrix representation:
\begin{align}
	\label{eqn:polarnondiag}
	\bold{\Pi}^{+}(\omega) = \bold{T}( \bold{\omega}\bold{1}-\bold{M}  )^{-1} {\bold{T}}^{\dagger}
\end{align}
Expanding the effective Hamiltonian ($\bold{M}$) and transition moments ($\bold{T}$) matrices in the M\o ller--Plesset perturbation series\cite{moller:1934p618} up to order $n$
\begin{align}
	\label{eqn:M_pt}
	\bold{M}  &\approx \bold{M}^{(0)} + \bold{M}^{(1)} + \ldots + \bold{M}^{(n)} \ ,  \\
	\label{eqn:T_pt}
	\bold{T} &\approx \bold{T}^{(0)} + \bold{T}^{(1)} + \ldots + \bold{T}^{(n)}\ 
\end{align}
defines the $\mathit{n}$th-order single-reference ADC method (ADC($n$)).
An alternative approach that constructs the ADC approximations for the polarization propagator using multireference perturbation theory has been described elsewhere.\cite{Sokolov:2018p204113,Mazin:2021p6152}

ADC allows to directly calculate excitation energies as eigenvalues of the effective Hamiltonian matrix $\bold{M}$
\begin{equation}
\label{eqn:eigenvalue}
\bold{M}\bold{Y}=\bold{Y}\bold{\Omega}
\end{equation}
The resulting eigenvectors $\bold{Y}$ are used to calculate the ADC($n$) spectroscopic amplitudes  $\bold{X}=\bold{T}\bold{Y}$, which are combined with transition dipole moment matrix elements ($\mu_{\alpha,pq}$, $\alpha = x, y, z$) to compute oscillator strengths 
\begin{align}
\label{eqn:oscillator}
f_{m} =\frac{2}{3}\Omega_{m}\sum_{\alpha=x,y,z}\left(\sum_{pq}\mu_{\alpha,pq}X_{pq,m}\right)^2
\end{align}
for the electronic transition with excitation energy $\Omega_{m}$ ($m > 0$). 

\begin{figure*}[t!]
	\centering
	\captionsetup{justification=raggedright,singlelinecheck=false,font=footnotesize}
	\includegraphics[scale=0.50,trim=0.0cm 0.0cm 0.0cm 0.0cm,clip]{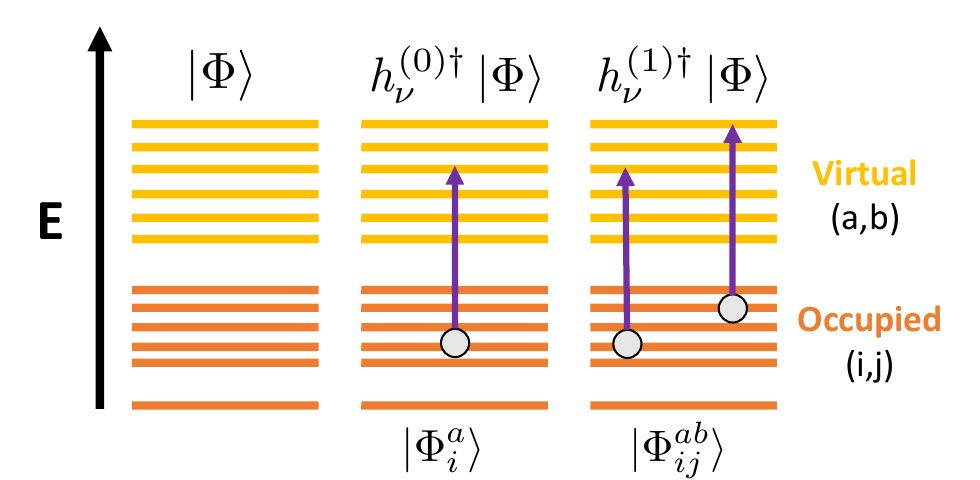}
	\caption{Electronic configurations $h_{\nu}^{(m)\dagger} \ket{\Phi}$ ($m=0,1$) used for representing the excited states in low-order ADC($n$) approximations ($n \le 3$), relative to the Hartree--Fock reference wavefunction $\ket{\Phi}$. }
	\label{fig:excitation_man}
\end{figure*}

Matrix elements of $\bold{M}$ derived using the effective Liouvillian approach\cite{Prasad:1985p1287,Mukherjee:1989p257,Sokolov:2018p204113,banerjee:2019p224112} for each perturbation order $(n)$ have the form:
\begin{align}
	\label{eqn:M_plus}
	M_{\mu\nu}^{(n)} &= \sum_{klm}^{k+l+m=n} \braket{\Phi|[h_{\mu}^{(k)},[\tilde{H}^{(l)},h_{\nu}^{(m)\dagger}]]|\Phi}
\end{align}
where $\ket{\Phi}$ denotes the reference Hartree--Fock wavefunction, $\tilde{H}^{(l)}$ is the $l$th-order contribution to the effective Hamiltonian (see Ref.\@ \citenum{Sokolov:2018p204113} for explicit expressions), and $h_{\nu}^{(m)\dagger}$ are the excitation operators that form the basis of electronic configurations for representing excited states.
In the low-order ADC($n$) approximations ($n \le 3$), only two types of $h_{\nu}^{(m)\dagger}$ need to be considered: the single-excitation operators $\mathit{h}_{\nu}^{(0)\dagger}=a_{a}^{\dagger}a_{i}$ and the double-excitation operators $\mathit{h}_{\nu}^{(1)\dagger}=a_{a}^{\dagger}a_{b}^{\dagger}a_{j}a_{i}$, where $i,j$ and $a,b$ denote occupied and virtual molecular orbitals, respectively. 
The resulting electronic configurations $h_{\nu}^{(m)\dagger} \ket{\Phi}$ are schematically depicted in \cref{fig:excitation_man}.
Similarly, the $n$th-order contributions to the matrix elements of $\bold{T}$ can be expressed as:
\begin{align}
	\label{eqn:T}
	T_{pq\nu}^{(n)} &= \sum_{kl}^{k+l=n} \braket{\Phi|[\tilde{q}_{pq}^{(k)}, h_{\nu}^{(l)\dagger}]|\Phi},
\end{align}
where $\tilde{q}_{pq}^{(k)}$ is the $k$th-order effective observable operator\cite{Sokolov:2018p204113}  with   $\tilde{q}_{pq}^{(0)} = \c{p} \a{q} - \braket{\Phi|\c{p} \a{q}|\Phi}$. 

\begin{figure*}[t!]
	\centering
	\captionsetup{justification=raggedright,singlelinecheck=false,font=footnotesize}
	\includegraphics[scale=0.40,trim=0.0cm 0.0cm 0.0cm 0.0cm,clip]{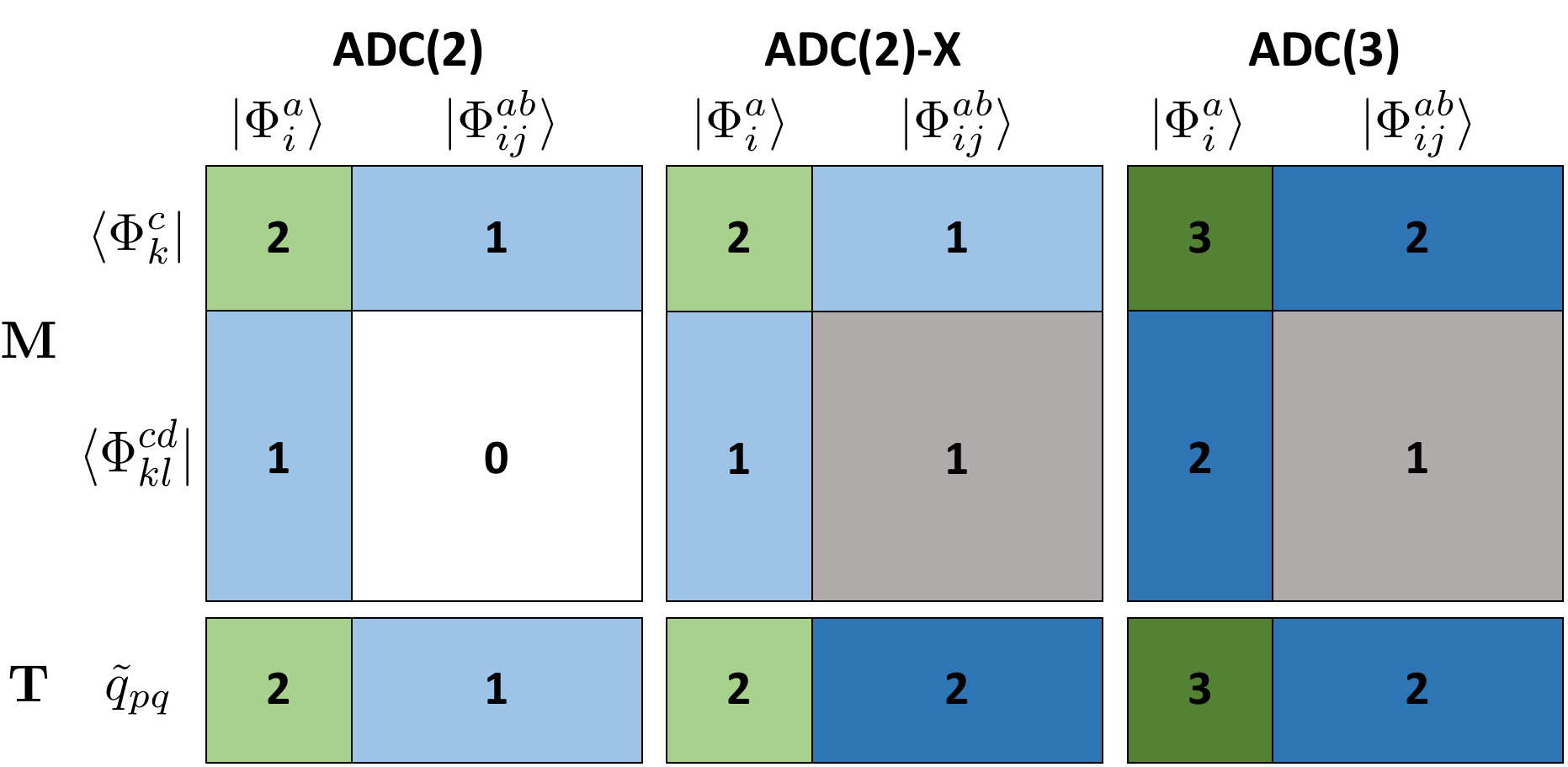}
	\caption{
	Schematic illustration of perturbative structure in the $\bold{M}$ and $\bold{T}$ matrices. 
	The wavefunctions $\ket{\Phi_{i}^{a}}$ and $\ket{\Phi_{ij}^{ab}}$ indicate single and double excitations relative to the Hartree--Fock reference, respectively (\cref{fig:excitation_man}).
	Numbers denote the perturbation order at which the expansions of effective Hamiltonian $\tilde{H}$ and observable $\tilde{q}_{pq}$ operators are truncated in each block.
	}
	\label{fig:m_matrices}
\end{figure*}
The $\bold{M}$ and $\bold{T}$ matrices have special perturbative structures that are depicted in \cref{fig:m_matrices} for the low-order ADC methods. 
In addition to the strict second-order (ADC(2)) and third-order (ADC(3)) approximations, in this work we will also consider the ``extended'' second-order method (ADC(2)-X), which incorporates third-order contributions to the effective Hamiltonian for the description of double excitations.

\subsection{Spin expectation values and contamination in ADC}
\label{sec:theory:Spin contamination}

The single-reference ADC calculations of open-shell molecules may suffer from spin symmetry breaking in (i) the reference Hartree--Fock (HF) wavefunction and/or (ii) the eigenvectors of ADC($n$) effective Hamiltonian matrix representing excited states.
The reference spin contamination (SC) originates from using the unrestricted HF (UHF) molecular orbitals and can be eliminated by employing the restricted open-shell HF (ROHF) reference instead.
The SC in ADC($n$) eigenvectors arises when the manifold of excitations described by the operators $h_{\nu}^{(m)\dagger}$  (\cref{fig:excitation_man}) is incomplete for representing a particular spin eigenstate. 
The spin symmetry breaking of this kind cannot be completely eliminated even when using the ROHF reference wavefunction, unless rigorous spin adaptation of equations is performed.

For an approximate wavefunction $\ket{\Psi}$ corresponding to an exact eigenstate with quantum number $S$, the SC in atomic units is defined as:\cite{stuck:2013p244109,andrews:1991p423,krylov:2017p151}
\begin{align}
	\label{eqn:spin_c}
	\Delta\mathit{S}^{2} &= \braket{\Psi|\hat{S}^{2}|\Psi}-S(S+1) \notag \\
	&\equiv \braket{{S}^{2}} - S(S+1)
\end{align}
where $\hat{S}^{2}$ is the spin-square operator.
Calculating $\Delta\mathit{S}^{2}$ requires evaluating the expectation value $\braket{\Psi|\hat{S}^{2}|\Psi}$, which can be expressed as\cite{kouba:1969p513,purvis:1988p2203,Stanton:1994p371,krylov:2000p6052} 
\begin{align}
	\label{eqn:spin_squared}
	\braket{S^{2}} &=  \frac{1}{4}(\sum_{p\in \alpha} \gamma_p^p - \sum_{\bar{p}\in \beta} \gamma_{\bar{p}}^{\bar{p}})^{2} \notag \\
	& + \frac{1}{2}(\sum_{p\in \alpha} \gamma_p^p + \sum_{\bar{p}\in \beta} \gamma_{\bar{p}}^{\bar{p}}) 
	+ \sum_{\substack{p,s\in\alpha \\ \bar{q},\bar{r}\in\beta}}S^{p}_{\bar{r}}S^{\bar{q}}_{s}\Gamma^{p\bar{q}}_{\bar{r}s} 
\end{align}
in terms of the overlap matrix ($S^p_{q} = \braket{\phi_p | \phi_q}$) and the one- and two-particle reduced density matrices ($\gamma^p_q = \braket{\Psi|\c{p}\a{q}|\Psi}$ and $\Gamma^{pq}_{rs} = \braket{\Psi|\c{p}\c{q}\a{s}\a{r}|\Psi}$, respectively). 
In \cref{eqn:spin_squared}, a bar is used to distinguish the spin-up and spin-down molecular orbitals from each other.

To compute $\braket{S^{2}}$ and $\Delta\mathit{S}^{2}$ in ADC calculations of electronic excitations, we derived the equations for $\gamma^p_q$ and $\Gamma^{pq}_{rs}$ using the formalism of effective Liouvillian theory.\cite{Prasad:1985p1287,Mukherjee:1989p257,Sokolov:2018p204113,banerjee:2019p224112,stahl2022quantifying}
Defining $K \equiv \c{p}\a{q}$ or $\c{p}\c{q}\a{s}\a{r}$, the density matrices for the $I$th excited state described by the ADC eigenvector $Y_{\nu I}$ are expressed as:\cite{stahl2022quantifying}
\begin{align}
	\label{eqn:op_exp_value_adc}
	\braket{K}_{I} = \sum_{(m)}^{(n)} \sum_{\mu\nu} Y^{\dag}_{I \mu} K^{(m)}_{\mu\nu} Y_{\nu I} \
\end{align}
where $(n)$ indicates the perturbation order of ADC($n$) approximation and $K^{(m)}_{\mu\nu}$ is the $m$th-order contribution to the matrix elements of effective density operator ($\tilde{K}$):
\begin{align}
	\label{eqn:O_plus}
	K^{(n)}_{\mu\nu} = \sum_{klm}^{k+l+m=n} \braket{\Phi|h_{\mu}^{(k)}\tilde{K}^{(l)}h_{\nu}^{(m)\dagger}|\Phi} \ .
\end{align}
The operators $\tilde{K}^{(l)}$ at each order $l$ are obtained from the perturbative analysis of Baker--Campbell--Hausdorff expansion of $\tilde{K}$, as described in Ref.\@ \citenum{stahl2022quantifying}.
Combining \cref{eqn:op_exp_value_adc,eqn:O_plus} allows to obtain the expressions for $\gamma^p_q$ and $\Gamma^{pq}_{rs}$, which are provided in the Supplementary Material and can be used to evaluate $\braket{S^{2}}$ and $\Delta\mathit{S}^{2}$ for any excited state in ADC calculations. 
We note that an alternative approach for calculating operator expectation values in ADC based on intermediate state representation was developed by Schirmer, Trofimov, Dempwolff, and co-workers.\cite{Schirmer:2004p11449,Trofimov:2005p144115,Knippenberg:2012p64107,plasser:2014p24106,plasser2014p24107,dempwolff:2020p024113,Dempwolff:2021p104117} 

\section{Implementation and computational details}
\label{sec:Imp/comp_details}

To carry out the calculations of ADC excitation energies, oscillator strengths, and spin contamination, we developed a new unrestricted ADC implementation in the PySCF package\cite{sun:2020p24109} that supports the UHF and ROHF reference wavefunctions.
The ROHF-based ADC methods incorporated the first-order single excitation amplitudes and off-diagonal elements of Fock matrix in their equations as described in Ref.\@ \citenum{stahl2022quantifying}.
We note that due to the unrestricted nature of our ADC implementation, starting with the spin-adapted ROHF reference may still result in significantly spin-contaminated electronic states as demonstrated in \cref{sec:results}. 
The calculations of spin expectation values and contamination were performed using the excited-state reduced density matrices discussed in \cref{sec:theory:Spin contamination}.
To reduce the cost of ADC(3) computations, two approximations were introduced: 1) third-order amplitudes were neglected in the effective transition moments matrix $\mathbf{T}$ and 2) the contributions from $(k,l,m)$ = $(0,3,0)$, $(1,2,0)$, and $(0,2,1)$ in \cref{eqn:O_plus} were omitted in the excited-state one- and two-particle reduced density matrices (i.e., the density matrices were computed at the ADC(2)-X level of theory).
No approximations were introduced for ADC(2) and ADC(2)-X. 
All ADC calculations were performed using density fitting,\cite{whitten:1973p4496,vahtras:1993p514,feyereisen:1993p359,dunlap:1979p3396,Weigend:1997p331,Weigend:1998p143,Hattig:2000p5154,Weigend:2002p4285,Hattig:2005p37} correlating all electrons in each molecule. 

To assess the effect of spin contamination on the accuracy of ADC(2), ADC(2)-X, and ADC(3) excitation energies (\cref{sec:results:energies}), we performed calculations for the benchmark set of open-shell molecules developed by Loos and co-workers.\cite{loos2020mountaineering}
The aug-cc-pVXZ (X = D, T) basis sets\cite{dunning:1989p1007,kendall:1992p6796,prascher:2011p69,woon:1993p1358} were employed in this study, with the cc-pVXZ-RI (X = D, T) auxiliary basis sets\cite{hattig:2005p59,weigend:2002p3175} used to approximate the two-electron integrals in density fitting.
The aug-cc-pVDZ results are provided in the Supplementary Material. 
The molecular geometries were taken from Ref.\@ \citenum{loos2020mountaineering}.

To correlate spin contamination with the errors in ADC oscillator strengths in \cref{sec:results:osc_strengths}, transition properties were computed using linear-response coupled cluster theory with single, double, and triple excitations and ROHF reference (LR-CCSDT/ROHF).
These calculations were performed using the aug-cc-pVDZ basis set and the MRCC package.\cite{kallay2020mrcc,mrcc} 

Finally, in the study of phenyl radical UV/Vis spectrum (\cref{sec:results:phenyl}), the equilibrium geometry was optimized at the MP2 level of theory with the aug-cc-pVTZ basis set, the UHF reference wavefunction, and the Q-chem software package.\cite{epifanovsky:2021p84801}
The ADC(2) and ADC(2)-X simulations of UV/Vis spectra were performed using the aug-cc-pVTZ basis set.
Due to the high computational cost, the ADC(3) calculations were performed using the modified aug-cc-pVTZ basis set where the f functions were removed for all carbon atoms.
Numerical tests on smaller systems demonstrated the high accuracy of this approximation.
The UV/Vis spectra were simulated by plotting a spectral function of the form:
\begin{equation}
	\label{eq:spectralfunction_plot}
	T(\om) = -\frac{1}{\pi} \operatorname{Im} \left[ \sum_{k} \frac{f_k}{\om - \Omega_k + i\mkern1mu \eta } \right]
\end{equation}
where $f_k$ and $\Omega_k$ are the oscillator strength (\cref{eqn:oscillator}) and energy of the $k$th transition, $\eta$ is a broadening parameter set to 500 \cm.

\section{Results and Discussion}
\label{sec:results}

\subsection{Spin contamination and errors in excitation energies}
\label{sec:results:energies}

	\begin{table*}[t!]
	\captionsetup{justification=raggedright,singlelinecheck=false,font=footnotesize}
	\caption{
	Spin contamination (\SC, a.u.) in the excited states of weakly spin-contaminated molecules computed using the ADC methods with the UHF or ROHF reference and the aug-cc-pVTZ basis set. 
	The second column reports the spin contamination in the reference UHF wavefunction. 
	}
	\label{tab:ADC_EE_s_2_tz_weak}
	\setlength{\extrarowheight}{2pt}
	\setstretch{1}
	\tiny
	\centering
	\hspace*{-0.8cm}
	\begin{threeparttable}
		\begin{tabular}{lccccccccc}
			\hline\hline
			Molecule        & Reference &                   Excitation                   & ADC(2) & ADC(2)-X & ADC(3)\tnote{a} & ADC(2) & ADC(2)-X & ADC(3)\tnote{a} &  \\
			                &    UHF    &                                                &  UHF   &   UHF    &  UHF   &  ROHF  &   ROHF   &      ROHF       &  \\ \hline
			\ce{BeF  }      &   0.00    &    ${X}{}^{2}\Sigma^{+}\rightarrow {1}^{2}\Pi$            &  0.00  &   0.00   &  0.00  &  0.00  &   0.00   &      0.00       &  \\
			                &           & ${X}{}^{2}\Sigma^{+}\rightarrow {1}^{2}\Sigma^{+}$        &  0.01  &   0.01   &  0.01  &  0.00  &   0.01   &      0.00       &  \\
			\ce{BeH  }      &   0.00    &    ${X}{}^{2}\Sigma^{+}\rightarrow {1}^{2}\Pi$            &  0.01  &   0.02   &  0.02  &  0.01  &   0.02   &      0.02       &  \\
			                &           &    ${X}{}^{2}\Sigma^{+}\rightarrow {2}^{2}\Pi$            &  0.03  &   0.03   &  0.02  &  0.01  &   0.03   &      0.02       &  \\
			\ce{BH2  }      &   0.00    &          $\tilde{X}{}^{2}A_{1} \rightarrow {1}^{2}B_{1}$  &  0.02  &   0.03   &  0.03  &  0.02  &   0.03   &      0.02       &  \\
			\ce{OH   }      &   0.01    &    ${X}{}^{2}\Pi\rightarrow {1}^{2}\Sigma^{+}$            &  0.03  &   0.04   &  0.04  &  0.03  &   0.03   &      0.03       &  \\
			\ce{F2BO }      &   0.01    &          $\tilde{X}{}^{2}B_{2} \rightarrow {1}^{2}B_{1}$  &  0.03  &   0.04   &  0.03  &  0.03  &   0.03   &      0.03       &  \\
			                &           &          $\tilde{X}{}^{2}B_{2} \rightarrow {1}^{2}A_{1}$  &  0.04  &   0.04   &  0.04  &  0.03  &   0.03   &      0.03       &  \\
			\ce{NH2  }      &   0.01    &          $\tilde{X}{}^{2}B_{1} \rightarrow {1}^{2}A_{1}$  &  0.04  &   0.05   &  0.04  &  0.03  &   0.03   &      0.03       &  \\
			\ce{CH3  }      &   0.01    &         $\tilde{X}{}^{2}A''_{2}\rightarrow {1}^{2}A'_{1}$ &  0.02  &   0.02   &  0.02  &  0.01  &   0.01   &      0.01       &  \\
			                &           &          $\tilde{X}{}^{2}A''_{2}\rightarrow {1}^{2}E'$    &  0.03  &   0.06   &  0.05  &  0.02  &   0.04   &      0.04       &  \\
			                &           &          $\tilde{X}{}^{2}A''_{2}\rightarrow {2}^{2}E'$    &  0.03  &   0.03   &  0.03  &  0.02  &   0.01   &      0.01       &  \\
			                &           &        $\tilde{X}{}^{2}A''_{2}\rightarrow {1}^{2}A''_{2}$ &  0.02  &   0.03   &  0.03  &  0.01  &   0.01   &      0.01       &  \\
			\ce{HOC  }      &   0.01    &             $\tilde{X}{}^{2}A'\rightarrow {1}^{2}A''$     &  0.04  &   0.05   &  0.04  &  0.03  &   0.03   &      0.02       &  \\
			\ce{F2BS }      &   0.01    &          $\tilde{X}{}^{2}B_{2} \rightarrow {1}^{2}B_{1}$  &  0.05  &   0.06   &  0.06  &  0.03  &   0.04   &      0.04       &  \\
			                &           &          $\tilde{X}{}^{2}B_{2} \rightarrow {1}^{2}A_{1}$  &  0.05  &   0.06   &  0.06  &  0.04  &   0.05   &      0.04       &  \\
			\ce{HCO  }      &   0.02    &             $\tilde{X}{}^{2}A'\rightarrow {1}^{2}A''$     &  0.08  &   0.09   &  0.06  &  0.09  &   0.09   &      0.05       &  \\
			                &           &             $\tilde{X}{}^{2}A'\rightarrow {1}^{2}A'$      &  0.13  &   0.12   &  0.10  &  0.14  &   0.12   &      0.08       &  \\
			\ce{H2BO }      &   0.02    &          $\tilde{X}{}^{2}B_{2} \rightarrow {1}^{2}B_{1}$  &  0.04  &   0.05   &  0.04  &  0.03  &   0.03   &      0.03       &  \\
			                &           &          $\tilde{X}{}^{2}B_{2} \rightarrow {1}^{2}A_{1}$  &  0.05  &   0.06   &  0.05  &  0.03  &   0.04   &      0.03       &  \\
			\ce{PH2  }      &   0.02    &          $\tilde{X}{}^{2}B_{1} \rightarrow {1}^{2}A_{1}$  &  0.04  &   0.06   &  0.05  &  0.03  &   0.04   &      0.04       &  \\
			\ce{H2PO }      &   0.04    &             $\tilde{X}{}^{2}A'\rightarrow {1}^{2}A''$     &  0.09  &   0.11   &  0.11  &  0.07  &   0.08   &      0.09       &  \\
			                &           &             $\tilde{X}{}^{2}A'\rightarrow {1}^{2}A'$      &  0.05  &   0.07   &  0.07  &  0.07  &   0.09   &      0.10       &  \\
			\ce{H2PS }      &   0.04    &             $\tilde{X}{}^{2}A'\rightarrow {1}^{2}A''$     &  0.08  &   0.10   &  0.09  &  0.04  &   0.05   &      0.05       &  \\
			                &           &             $\tilde{X}{}^{2}A'\rightarrow {1}^{2}A'$      &  0.06  &   0.08   &  0.08  &  0.06  &   0.08   &      0.07       &  \\
			\ce{NO   }      &   0.04    &    ${X}{}^{2}\Pi \rightarrow {1}^{2}\Sigma^{+}$           &  0.13  &   0.13   &  0.09  &  0.09  &   0.09   &      0.06       &  \\
			                &           &        ${X}{}^{2}\Pi \rightarrow {2}^{2}\Sigma^{+}$       &  0.13  &   0.12   &  0.08  &  0.09  &   0.08   &      0.05       &  \\ \hline
			Sum             &           &                                                &  1.34  &   1.55   &  1.32  &  1.05  &   1.21   &      1.01       &  \\
			Average         &           &                                                &  0.05  &   0.06   &  0.05  &  0.04  &   0.04   &      0.04       &  \\
			Minimum &           &                                                &  0.00  &   0.00   &  0.00  &  0.00  &   0.00   &      0.00       &  \\
			Maximum &           &                                                &  0.13  &   0.13   &  0.11  &  0.14  &   0.12   &      0.10       &  \\ \hline\hline
		\end{tabular}
		\begin{tablenotes}
			\item[a] \scriptsize{ADC(3) \SC was computed using the ADC(2)-X density matrices and ADC(3) eigenvectors of effective Hamiltonian matrix (\cref{sec:Imp/comp_details}).}
		\end{tablenotes}
	\end{threeparttable}
\end{table*}	

	\begin{table*}[t!]
	\captionsetup{justification=raggedright,singlelinecheck=false,font=footnotesize}
	\caption{
	Spin contamination (\SC, a.u.) in the excited states of strongly spin-contaminated molecules computed using the ADC methods with the UHF or ROHF reference and the aug-cc-pVTZ basis set. 
	The second column reports the spin contamination in the reference UHF wavefunction. 
	}
	\label{tab:ADC_EE_s_2_tz_strong}
	\setlength{\extrarowheight}{2pt}
	\setstretch{1}
	\tiny
	\centering
	\hspace*{-0.8cm}
	\begin{threeparttable}
		\begin{tabular}{lccccccccc}
			\hline\hline
			Molecule         & Reference &                   Excitation                   & ADC(2) & ADC(2)-X & ADC(3)\tnote{a} & ADC(2) & ADC(2)-X & ADC(3)\tnote{a} &  \\
			                 &    UHF    &                                                &  UHF   &   UHF    &       UHF       &  ROHF  &   ROHF   &      ROHF       &  \\ \hline
			\ce{CNO  }       &   0.06    & ${X}{}^{2}\Pi \rightarrow {1}^{2}\Sigma^{+}$                                                                                                                &  0.25  &   0.21   &      0.29       &  0.29  &   0.24   &      0.33       &  \\
			                 &           &  ${X}{}^{2}\Pi \rightarrow {1}^{2}\Pi$   [$(1\pi)^{4}(2\pi_{x})^{2}(2\pi_y)^{1}\rightarrow(1\pi_{x})^{2}(1\pi_{y})^{1}(2\pi)^{4}$]                          &  0.12  &   0.12   &      0.24       &  0.14  &   0.49   &      0.13       &  \\
			                 &           &  ${X}{}^{2}\Pi \rightarrow {1}^{2}\Pi$   [$(1\pi)^{4}(2\pi_{x})^{2}(2\pi_y)^{1}\rightarrow(1\pi_{x})^{1}(1\pi_{y})^{2}(2\pi)^{4}$]                          &  0.60  &   0.30   &      0.62       &  0.47  &  0.35    &      0.41       &  \\
			\ce{NCO  }       &   0.09    & ${X}{}^{2}\Pi \rightarrow {1}^{2}\Sigma^{+}$                                                                                                                &  0.13  &   0.15   &      0.14       &  0.05  &   0.06   &      0.05       &  \\
			                 &           &  ${X}{}^{2}\Pi \rightarrow {1}^{2}\Pi$   [$(1\pi)^{4}(2\pi_{x})^{2}(2\pi_y)^{1}\rightarrow(1\pi_{x})^{2}(1\pi_{y})^{1}(2\pi)^{4}$]                          &  0.14  &   0.15   &      0.18       &  0.17  &   0.20   &      0.28       &  \\
			                 &           &  ${X}{}^{2}\Pi \rightarrow {1}^{2}\Pi$   [$(1\pi)^{4}(2\pi_{x})^{2}(2\pi_y)^{1}\rightarrow(1\pi_{x})^{1}(1\pi_{y})^{2}(2\pi)^{4}$]                          &  0.26  &   0.21   &      0.32       &  0.15  &   0.15   &      0.26       &  \\
			\ce{Vinyl}       &   0.20    &            $\tilde{X}{}^{2}A'\rightarrow {1}^{2}A''$                                                                                                        &  0.07  &   0.06   &      0.07       &  0.05  &   0.05   &      0.05       &  \\
			                 &           &            $\tilde{X}{}^{2}A'\rightarrow {2}^{2}A''$                                                                                                        &  0.11  &   0.11   &      0.11       &  0.07  &   0.08   &      0.07       &  \\
			                 &           &            $\tilde{X}{}^{2}A'\rightarrow {1}^{2}A'$                                                                                                         &  0.24  &   0.26   &      0.25       &  0.08  &   0.10   &      0.09       &  \\
			                 &           &            $\tilde{X}{}^{2}A'\rightarrow {2}^{2}A'$                                                                                                         &  0.25  &   0.26   &      0.26       &  0.09  &   0.10   &      0.10       &  \\
			\ce{Allyl}       &   0.20    &         $\tilde{X}{}^{2}A_{2} \rightarrow {1}^{2}B_{1}$                                                                                                     &  0.04  &   0.05   &      0.05       &  0.03  &   0.03   &      0.03       &  \\
			                 &           &         $\tilde{X}{}^{2}A_{2} \rightarrow {1}^{2}A_{1}$                                                                                                     &  0.21  &   0.23   &      0.22       &  0.03  &   0.05   &      0.04       &  \\
			\ce{CO+  }       &   0.22    &        ${X}{}^{2}\Sigma^{+}\rightarrow {1}^{2}\Pi$                                                                                                          &  0.13  &   0.15   &      0.15       &  0.04  &   0.05   &      0.07       &  \\
			                 &           &    ${X}{}^{2}\Sigma^{+}\rightarrow {1}^{2}\Sigma^{+}$                                                                                                       &  0.28  &   0.27   &      0.40       &  0.06  &   0.08   &      0.52       &  \\
			\ce{CH   }       &   0.35    &         ${X}{}^{2}\Pi \rightarrow {1}^{2}\Delta$ [$(3\sigma)^{2}(1\pi_y)^{1} \rightarrow (3\sigma)^{1}(1\pi_{x})^{1}(1\pi_y)^{1}$]                          &  0.08  &   0.48   &      0.04       &  0.02  &   0.01   &      0.01       &  \\
			                 &           &         ${X}{}^{2}\Pi \rightarrow {1}^{2}\Delta$  [$(3\sigma)^{2}(1\pi_y)^{1} \rightarrow(3\sigma)^{1}(1\pi_y)^{2}$]                                        &  0.04  &   0.05   &      0.67       &  0.02  &   0.30   &      0.10       &  \\
			\ce{CN   }       &   0.40    &        ${X}{}^{2}\Sigma^{+}\rightarrow {1}^{2}\Pi$                                                                                                          &  0.20  &   0.21   &      0.22       &  0.04  &   0.05   &      0.06       &  \\
			                 &           &    ${X}{}^{2}\Sigma^{+}\rightarrow {1}^{2}\Sigma^{+}$                                                                                                       &  0.39  &   0.37   &      0.40       &  0.05  &   0.07   &      0.15       &  \\
			\ce{Nitromethyl} &   0.42    &         $\tilde{X}{}^{2}B_{1} \rightarrow {1}^{2}B_{2}$                                                                                                     &  0.61  &   0.51   &      0.65       &  0.30  &   0.30   &      0.49       &  \\
			                 &           &         $\tilde{X}{}^{2}B_{1} \rightarrow {1}^{2}A_{2}$                                                                                                     &  0.29  &   0.21   &      0.27       &  0.61  &   0.45   &      0.65       &  \\
			                 &           &         $\tilde{X}{}^{2}B_{1} \rightarrow {1}^{2}A_{1}$                                                                                                     &  0.54  &   0.58   &      0.53       &  0.26  &   0.27   &      0.40       &  \\
			                 &           &         $\tilde{X}{}^{2}B_{1} \rightarrow {1}^{2}B_{1}$                                                                                                     &  0.30  &   0.21   &      0.43       &  0.13  &   0.17   &      0.16       &  \\
			\ce{CON  }       &   0.68    &           ${X}{}^{2}\Pi \rightarrow {1}^{2}\Pi$  [$(1\pi)^{4}(2\pi_{x})^{2}(2\pi_y)^{1}\rightarrow (1\pi)^{4}(2\pi_{x})^{1}(2\pi_y)^{1}(3\pi_{x})^{1}$]     &  0.56  &   0.88   &      0.92       &  0.33    &   0.56   &      0.11       &  \\
			                 &           &           ${X}{}^{2}\Pi \rightarrow {1}^{2}\Pi$ [$(1\pi)^{4}(2\pi_{x})^{2}(2\pi_y)^{1}\rightarrow (1\pi)^{4}(2\pi_{x})^{1}(2\pi_y)^{1}(3\pi_{y})^{1}$]      &  0.78  &   0.49   &      0.57       & 0.02 &   0.06   &   0.68          &  \\ \hline
			Sum              &           &                                                &  6.61  &   6.50   &      7.99       &  3.51  &   4.26   &      5.26       &  \\
			Average          &           &                                                &  0.28  &   0.27   &      0.33       &  0.15  &   0.18   &      0.22       &  \\
			Minimum          &           &                                                &  0.04  &   0.05   &      0.04       &  0.02  &   0.01   &      0.01       &  \\
			Maximum          &           &                                                &  0.78  &   0.88   &      0.92       &  0.61  &   0.56   &      0.68       &  \\ \hline\hline
		\end{tabular}
		\begin{tablenotes}
			\item[a] \scriptsize{ADC(3) \SC was computed using the ADC(2)-X density matrices and ADC(3) eigenvectors of effective Hamiltonian matrix (\cref{sec:Imp/comp_details}).}
		\end{tablenotes}
	\end{threeparttable}
\end{table*}

        \begin{table*}[t!]
		\captionsetup{justification=raggedright,singlelinecheck=false,font=footnotesize}
        \caption{
		Vertical excitation energies (eV) of weakly spin-contaminated molecules computed using the ADC methods with the UHF or ROHF reference and the aug-cc-pVTZ basis set. 
		For each method, we report the mean absolute errors (MAE)  and standard deviations of errors (STDV) relative to the full configuration interaction (FCI) results from Ref.\@ \citenum{loos2020mountaineering}. 
		Minimum (MIN) and maximum (MAX) errors are also shown. 
        }
        \label{tab:ADC_EE_TZ_energies_weak}
        \setlength{\extrarowheight}{2pt}
        \setstretch{1}
        \tiny
        \centering
        \hspace*{-0.8cm}
        \begin{threeparttable}
                \begin{tabular}{lccccccccc}
                	\hline\hline
                	Molecule   &                       Excitation                        & ADC(2) & ADC(2)-X & ADC(3) & ADC(2) & ADC(2)-X & ADC(3) &      FCI      &  \\
                	           &                                                         &  UHF   &   UHF    &  UHF   &  ROHF  &   ROHF   &  ROHF  &               &  \\ \hline
                	\ce{BeF  } &    ${X}{}^{2}\Sigma^{+}\rightarrow {1}^{2}\Pi$            &  4.21  &   4.12   &  4.11  &  4.20  &   4.11   &  4.11  &     4.14      &  \\
                	           & ${X}{}^{2}\Sigma^{+}\rightarrow {1}^{2}\Sigma^{+}$        &  6.36  &   6.24   &  6.23  &  6.35  &   6.24   &  6.23  &     6.21      &  \\
                	\ce{BeH  } &    ${X}{}^{2}\Sigma^{+}\rightarrow {1}^{2}\Pi$            &  2.60  &   2.35   &  2.44  &  2.57  &   2.33   &  2.44  &     2.49      &  \\
                	           &    ${X}{}^{2}\Sigma^{+}\rightarrow {2}^{2}\Pi$            &  6.54  &   6.34   &  6.43  &  6.53  &   6.34   &  6.43  &     6.46      &  \\
                	\ce{BH2  } &          $\tilde{X}{}^{2}A_{1} \rightarrow {1}^{2}B_{1}$  &  1.31  &   0.88   &  1.11  &  1.29  &   0.86   &  1.11  &     1.18      &  \\
                	\ce{OH   } &    ${X}{}^{2}\Pi\rightarrow {1}^{2}\Sigma^{+}$            &  4.28  &   3.66   &  4.04  &  4.22  &   3.62   &  4.02  &     4.10      &  \\
                	\ce{F2BO } &          $\tilde{X}{}^{2}B_{2} \rightarrow {1}^{2}B_{1}$  &  0.82  &   0.30   &  0.59  &  0.79  &   0.27   &  0.58  &     0.71      &  \\
                	           &          $\tilde{X}{}^{2}B_{2} \rightarrow {1}^{2}A_{1}$  &  2.92  &   2.32   &  2.73  &  2.88  &   2.29   &  2.72  &     2.78      &  \\
                	\ce{NH2  } &          $\tilde{X}{}^{2}B_{1} \rightarrow {1}^{2}A_{1}$  &  2.22  &   1.65   &  2.04  &  2.18  &   1.61   &  2.02  &     2.12      &  \\
                	\ce{CH3  } &         $\tilde{X}{}^{2}A''_{2}\rightarrow {1}^{2}A'_{1}$ &  6.02  &   5.49   &  5.83  &  5.96  &   5.44   &  5.81  &     5.85      &  \\
                	           &          $\tilde{X}{}^{2}A''_{2}\rightarrow {1}^{2}E'$    &  7.19  &   6.49   &  6.92  &  7.12  &   6.43   &  6.89  &     6.96      &  \\
                	           &          $\tilde{X}{}^{2}A''_{2}\rightarrow {2}^{2}E'$    &  7.34  &   6.80   &  7.17  &  7.27  &   6.76   &  7.15  &     7.18      &  \\
                	           &        $\tilde{X}{}^{2}A''_{2}\rightarrow {1}^{2}A''_{2}$ &  7.76  &   7.29   &  7.62  &  7.71  &   7.26   &  7.60  &     7.65      &  \\
                	\ce{HOC  } &             $\tilde{X}{}^{2}A'\rightarrow {1}^{2}A''$     &  1.03  &   0.37   &  0.71  &  0.97  &   0.32   &  0.69  &     0.92      &  \\
                	\ce{F2BS } &          $\tilde{X}{}^{2}B_{2} \rightarrow {1}^{2}B_{1}$  &  0.63  &  -0.03   &  0.35  &  0.57  &  -0.07   &  0.32  &     0.48      &  \\
                	           &          $\tilde{X}{}^{2}B_{2} \rightarrow {1}^{2}A_{1}$  &  3.19  &   2.39   &  2.88  &  3.11  &   2.33   &  2.84  &     2.93      &  \\
                	\ce{HCO  } &             $\tilde{X}{}^{2}A'\rightarrow {1}^{2}A''$     &  2.24  &   1.42   &  1.92  &  2.19  &   1.41   &  1.92  &     2.09      &  \\
                	           &             $\tilde{X}{}^{2}A'\rightarrow {1}^{2}A'$      &  5.38  &   4.85   &  5.53  &  5.38  &   4.86   &  5.51  &     5.42      &  \\
                	\ce{H2BO } &          $\tilde{X}{}^{2}B_{2} \rightarrow {1}^{2}B_{1}$  &  2.10  &   1.58   &  1.81  &  2.01  &   1.49   &  1.76  &     2.15      &  \\
                	           &          $\tilde{X}{}^{2}B_{2} \rightarrow {1}^{2}A_{1}$  &  3.53  &   2.88   &  3.28  &  3.45  &   2.80   &  3.23  &     3.49      &  \\
                	\ce{PH2  } &          $\tilde{X}{}^{2}B_{1} \rightarrow {1}^{2}A_{1}$  &  2.98  &   2.22   &  2.65  &  2.88  &   2.17   &  2.62  &     2.77      &  \\
                	\ce{H2PO } &             $\tilde{X}{}^{2}A'\rightarrow {1}^{2}A''$     &  3.02  &   2.26   &  2.75  &  2.82  &   2.03   &  2.83  &     2.80      &  \\
                	           &             $\tilde{X}{}^{2}A'\rightarrow {1}^{2}A'$      &  4.42  &   3.73   &  3.93  &  4.19  &   3.42   &  3.96  &     4.19      &  \\
                	\ce{H2PS } &             $\tilde{X}{}^{2}A'\rightarrow {1}^{2}A''$     &  1.29  &   0.57   &  0.97  &  1.17  &   0.47   &  0.89  &     1.16      &  \\
                	           &             $\tilde{X}{}^{2}A'\rightarrow {1}^{2}A'$      &  2.86  &   2.05   &  2.49  &  2.75  &   1.88   &  2.39  &     2.72      &  \\
                	\ce{NO   } &    ${X}{}^{2}\Pi \rightarrow {1}^{2}\Sigma^{+}$           &  6.17  &   5.50   &  6.22  &  6.18  &   5.52   &  6.11  &     6.13      &  \\
                	           &        ${X}{}^{2}\Pi \rightarrow {2}^{2}\Sigma^{+}$       &  7.21  &   6.65   &  7.34  &  7.23  &   6.64   &  7.35  & 7.29\tnote{a} &  \\ \hline
                	\mae       &                                                         &  0.13  &   0.45   &  0.10  &  0.08  &   0.50   &  0.12  &               &  \\
                	\std       &                                                         &  0.08  &   0.19   &  0.11  &  0.07  &   0.22   &  0.12  &               &  \\
                	MIN        &                                                         &  0.04  &   0.02   &  0.01  &  0.00  &   0.03   &  0.02  &               &  \\
                	MAX        &                                                         &  0.26  &   0.67   &  0.34  &  0.18  &   0.84   &  0.39  &               &  \\ \hline\hline
                \end{tabular}
                \begin{tablenotes}
                        \item[{a}] Computed using equation-of-motion coupled cluster theory with up to quadruple excitations (EOM-CCSDTQ)\cite{loos2020mountaineering}
                \end{tablenotes}
        \end{threeparttable}
        \end{table*}

        \begin{table*}[t!]
		\captionsetup{justification=raggedright,singlelinecheck=false,font=footnotesize}
        \caption{
		Vertical excitation energies (eV) of strongly spin-contaminated molecules computed using the ADC methods with the UHF or ROHF reference and the aug-cc-pVTZ basis set. 
		For each method, we report the mean absolute errors (MAE)  and standard deviations of errors (STDV) relative to the full configuration interaction (FCI) results from Ref.\@ \citenum{loos2020mountaineering}. 
		Minimum (MIN) and maximum (MAX) errors are also shown. 
        }
        \label{tab:ADC_EE_TZ_energies_strong}
        \setlength{\extrarowheight}{2pt}
        \setstretch{1}
        \tiny
        \centering
        \hspace*{-0.8cm}
        \begin{threeparttable}
                \begin{tabular}{lccccccccc}
                	\hline\hline
                	Molecule         &                      Excitation                       & ADC(2) & ADC(2)-X & ADC(3) & ADC(2) & ADC(2)-X & ADC(3) &      FCI      &  \\
                	                 &                                                      &  UHF   &   UHF    &  UHF   &  ROHF  &   ROHF   &  ROHF  &               &  \\ \hline
                	\ce{CNO  }       & ${X}{}^{2}\Pi \rightarrow {1}^{2}\Sigma^{+}$                                                                                                                 &  2.45  &   0.64   &  1.95  &  2.47  &   0.45   &  1.75  &     1.61      &  \\
                	                 &  ${X}{}^{2}\Pi \rightarrow {1}^{2}\Pi$   [$(1\pi)^{4}(2\pi_{x})^{2}(2\pi_y)^{1}\rightarrow(1\pi_{x})^{2}(1\pi_{y})^{1}(2\pi)^{4}$]                           &  5.70  &   4.46   &  5.29  &  5.58  &   4.39   &  5.02  & 5.57\tnote{a} &  \\
                	                 &  ${X}{}^{2}\Pi \rightarrow {1}^{2}\Pi$   [$(1\pi)^{4}(2\pi_{x})^{2}(2\pi_y)^{1}\rightarrow(1\pi_{x})^{1}(1\pi_{y})^{2}(2\pi)^{4}$]                           &  5.99  &   4.54   &  5.59  &  5.83  &   4.06   &  5.17  & 5.57\tnote{a} &  \\
                	\ce{NCO  }       & ${X}{}^{2}\Pi \rightarrow {1}^{2}\Sigma^{+}$                                                                                                                 &  3.03  &   2.18   &  2.97  &  2.76  &   2.00   &  2.73  &     2.83      &  \\
                	                 &  ${X}{}^{2}\Pi \rightarrow {1}^{2}\Pi$   [$(1\pi)^{4}(2\pi_{x})^{2}(2\pi_y)^{1}\rightarrow(1\pi_{x})^{2}(1\pi_{y})^{1}(2\pi)^{4}$]                           &  5.22  &   3.86   &  4.64  &  4.84  &   3.34   &  4.60  &     4.70      &  \\
                	                 &  ${X}{}^{2}\Pi \rightarrow {1}^{2}\Pi$   [$(1\pi)^{4}(2\pi_{x})^{2}(2\pi_y)^{1}\rightarrow(1\pi_{x})^{1}(1\pi_{y})^{2}(2\pi)^{4}$]                           &  5.55  &   4.18   &  5.07  &  5.09  &   3.60   &  5.01  &     4.70      &  \\
                	\ce{Vinyl}       &            $\tilde{X}{}^{2}A'\rightarrow {1}^{2}A''$                                                                                                         &  4.05  &   2.53   &  3.15  &  3.54  &   2.62   &  3.04  &     3.26      &  \\
                	                 &            $\tilde{X}{}^{2}A'\rightarrow {2}^{2}A''$                                                                                                         &  5.20  &   4.01   &  4.66  &  4.93  &   4.05   &  4.59  &     4.69      &  \\
                	                 &            $\tilde{X}{}^{2}A'\rightarrow {1}^{2}A'$                                                                                                          &  6.64  &   5.81   &  6.39  &  6.45  &   5.59   &  6.23  &     5.60      &  \\
                	                 &            $\tilde{X}{}^{2}A'\rightarrow {2}^{2}A'$                                                                                                          &  7.23  &   6.39   &  6.98  &  7.03  &   6.18   &  6.82  & 6.21\tnote{a} &  \\
                	\ce{Allyl}       &         $\tilde{X}{}^{2}A_{2} \rightarrow {1}^{2}B_{1}$                                                                                                      &  4.34  &   2.64   &  3.38  &  3.77  &   2.61   &  3.12  & 3.43\tnote{a} &  \\
                	                 &         $\tilde{X}{}^{2}A_{2} \rightarrow {1}^{2}A_{1}$                                                                                                      &  5.34  &   4.59   &  5.05  &  5.01  &   4.31   &  4.79  & 4.97\tnote{a} &  \\
                	\ce{CO+  }       &        ${X}{}^{2}\Sigma^{+}\rightarrow {1}^{2}\Pi$                                                                                                           &  3.74  &   1.98   &  3.78  &  3.16  &   2.06   &  3.56  &     3.28      &  \\
                	                 &    ${X}{}^{2}\Sigma^{+}\rightarrow {1}^{2}\Sigma^{+}$                                                                                                        &  5.83  &   4.38   &  6.84  &  4.80  &   3.98   &  6.98  &     5.81      &  \\
                	\ce{CH   }       &         ${X}{}^{2}\Pi \rightarrow {1}^{2}\Delta$ [$(3\sigma)^{2}(1\pi_y)^{1} \rightarrow (3\sigma)^{1}(1\pi_{x})^{1}(1\pi_y)^{1}$]                           &  3.17  &   2.14   &  2.33  &  3.16  &   1.93   &  2.06  &     2.91      &  \\
                	                 &         ${X}{}^{2}\Pi \rightarrow {1}^{2}\Delta$  [$(3\sigma)^{2}(1\pi_y)^{1} \rightarrow(3\sigma)^{1}(1\pi_y)^{2}$]                                         &  3.75  &   2.18   &  2.37  &  3.42  &   2.72   &  3.02  &     2.91      &  \\
                	\ce{CN   }       &        ${X}{}^{2}\Sigma^{+}\rightarrow {1}^{2}\Pi$                                                                                                           &  2.06  &   0.12   &  1.09  &  1.41  &   0.38   &  0.89  &     1.34      &  \\
                	                 &    ${X}{}^{2}\Sigma^{+}\rightarrow {1}^{2}\Sigma^{+}$                                                                                                        &  3.71  &   1.78   &  3.41  &  2.53  &   1.38   &  3.21  &     3.22      &  \\
                	\ce{Nitromethyl} &         $\tilde{X}{}^{2}B_{1} \rightarrow {1}^{2}B_{2}$                                                                                                      &  2.23  &   0.91   &  2.03  &  2.01  &   0.92   &  2.33  & 2.05\tnote{a} &  \\
                	                 &         $\tilde{X}{}^{2}B_{1} \rightarrow {1}^{2}A_{2}$                                                                                                      &  2.60  &   0.58   &  2.01  &  2.83  &   1.19   &  1.96  & 2.38\tnote{a} &  \\
                	                 &         $\tilde{X}{}^{2}B_{1} \rightarrow {1}^{2}A_{1}$                                                                                                      &  2.82  &   1.72   &  2.98  &  2.56  &   1.47   &  2.70  & 2.56\tnote{a} &  \\
                	                 &         $\tilde{X}{}^{2}B_{1} \rightarrow {1}^{2}B_{1}$                                                                                                      &  6.06  &   4.45   &  5.21  &  5.78  &   4.50   &  5.19  & 5.35\tnote{a} &  \\
                	\ce{CON  }       &           ${X}{}^{2}\Pi \rightarrow {1}^{2}\Pi$  [$(1\pi)^{4}(2\pi_{x})^{2}(2\pi_y)^{1}\rightarrow (1\pi)^{4}(2\pi_{x})^{1}(2\pi_y)^{1}(3\pi_{x})^{1}$]      &  2.91  &   2.02   &  2.24  &  3.89  &   2.18   &  3.66  & 3.54\tnote{a} &  \\
                	                 &           ${X}{}^{2}\Pi \rightarrow {1}^{2}\Pi$ [$(1\pi)^{4}(2\pi_{x})^{2}(2\pi_y)^{1}\rightarrow (1\pi)^{4}(2\pi_{x})^{1}(2\pi_y)^{1}(3\pi_{y})^{1}$]       &  3.39  &   2.11   &  2.82  &  3.37  &   2.27   &  3.46  & 3.54\tnote{a} &  \\ \hline
                	\mae             &                                                       &  0.52  &   0.94   &  0.38  &  0.35  &   0.99   &  0.32  &               &  \\
                	\std             &                                                       &  0.40  &   0.49   &  0.52  &  0.43  &   0.48   &  0.43  &               &  \\
                	MIN              &                                                       &  0.02  &   0.18   &  0.02  &  0.00  &   0.01   &  0.01  &               &  \\
                	MAX              &                                                       &  1.04  &   1.80   &  1.30  &  1.01  &   1.84   &  1.17  &               &  \\ \hline\hline
                \end{tabular}
                \begin{tablenotes}
                        \item[{a}] Computed using equation-of-motion coupled cluster theory with up to triple excitations (EOM-CCSDT)\cite{loos2020mountaineering}
                \end{tablenotes}
        \end{threeparttable}
        \end{table*}

We begin by investigating the effect of ground- and excited-state spin contamination (\SC) on the accuracy of ADC excitation energies. 
We conduct our study for the benchmark set of Loos and co-workers,\cite{loos2020mountaineering} which contains highly accurate reference excitation energies for 24 open-shell molecules.
To organize our discussion, we split the molecules in this set into two groups: i) weakly spin-contaminated (WSM) with the UHF \SC $\le$ 0.05 (15 molecules) and ii) strongly spin-contaminated (SSM) with the UHF \SC $>$ 0.05 (9 molecules). 

\cref{tab:ADC_EE_s_2_tz_weak,tab:ADC_EE_s_2_tz_strong} show the excited-state \SC for WSM and SSM, respectively, computed using ADC(2), ADC(2)-X, and ADC(3) starting with either the UHF or ROHF reference wavefunction. 
As the ground-state UHF \SC increases, the ADC methods tend to exhibit higher spin contamination in the excited states.
This can be observed in the total, average, and maximum \SC that are significantly higher for SSM compared to WSM.
While for WSM \SC is relatively insensitive to the choice of reference wavefunction, using the ROHF reference in ADC calculations helps to significantly reduce \SC for SSM. 

Contrary to what might be expected, ADC(3) tends to show higher spin contamination in its excited states than ADC(2) and ADC(2)-X, which suggests that increasing the level of theory may worsen the description of electronic spin in the unrestricted ADC calculations for SSM. 
This is particularly noticeable for the ROHF-based ADC methods where the total and average \SC of SSM increases by $\sim$ 47 \% from ADC(2) to ADC(3).
We note that this trend may be affected by the approximations in ADC(3) calculations of \SC, which employed  the ADC(2)-X reduced density matrices and ADC(3) eigenvectors of effective Hamiltonian matrix (\cref{sec:Imp/comp_details}). 
However, an increase in \SC from ADC(2) to ADC(3) was also observed in the ADC calculations of charged excitations where no approximations in the ADC(3) density matrices were introduced.\cite{stahl2022quantifying}

\begin{figure*}[t!]
	\centering
	\captionsetup{justification=raggedright,singlelinecheck=false,font=footnotesize}
	\includegraphics[scale=0.46,trim=0.0cm 0.0cm 0.0cm 0.0cm,clip]{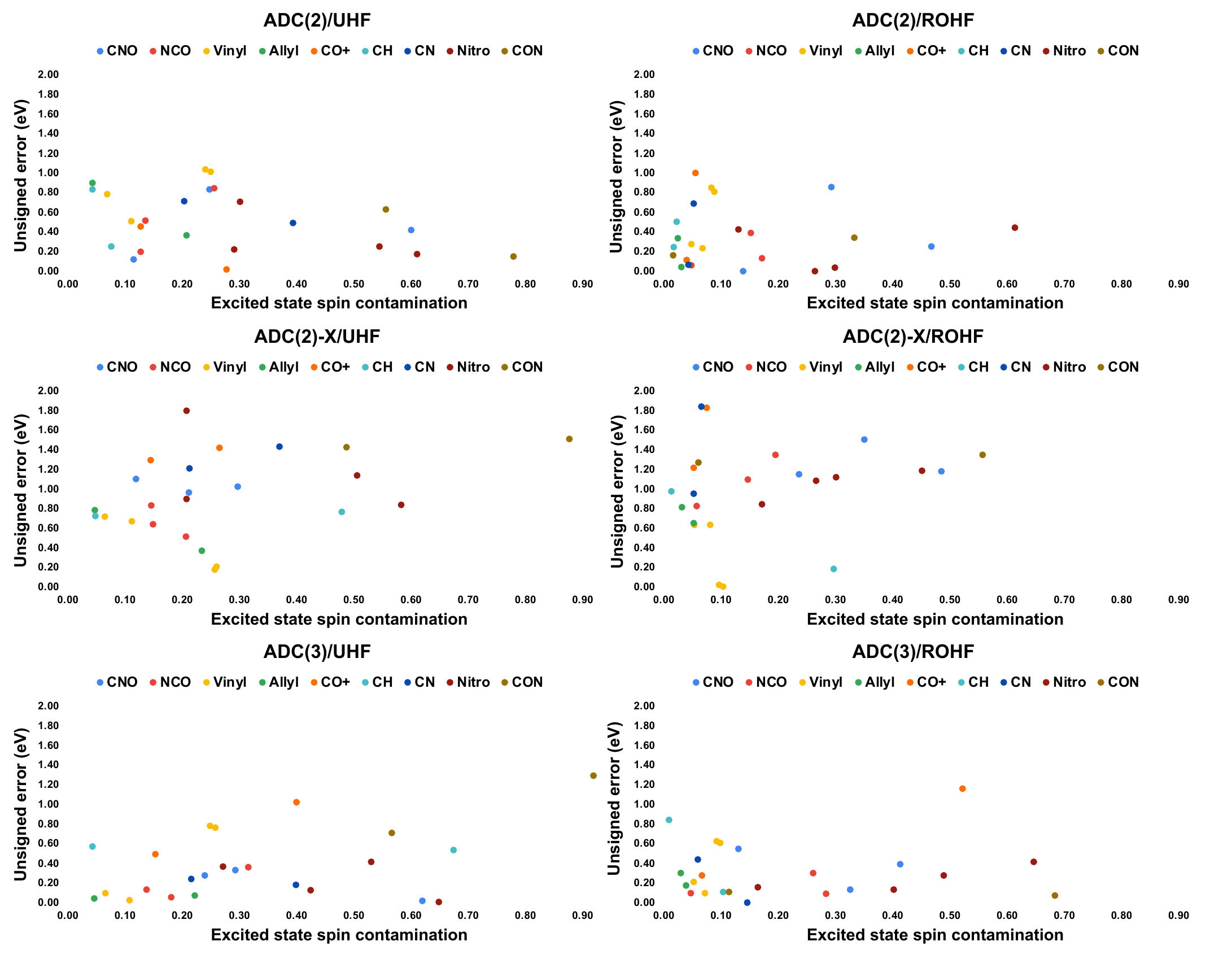}
	\caption{
		Unsigned errors in the excitation energies of strongly spin-contaminated molecules plotted against the excited-state spin contamination computed using the ADC methods with the UHF or ROHF reference wavefunctions.
		See \cref{tab:ADC_EE_TZ_energies_strong} for details.
	}
	\label{fig:ee_error_vs_sc}
\end{figure*}

To analyze the effect of \SC on the accuracy of ADC methods, we computed the errors in ADC(2), ADC(2)-X, and ADC(3) excitation energies relative to the accurate reference data from Loos et al.\cite{loos2020mountaineering}
These results, along with their statistical analysis, are shown in \cref{tab:ADC_EE_TZ_energies_weak,tab:ADC_EE_TZ_energies_strong} for WSM and SSM, respectively.
In addition, in \cref{fig:ee_error_vs_sc} the absolute errors in excitation energies are plotted against \SC for the excited states of SSM.
For WSM, the ADC(2) and ADC(3) methods show accurate results with either the UHF or ROHF reference wavefunctions.
This is reflected by their mean absolute errors (\mae) and standard deviation of errors (\std) that do not exceed 0.13 eV.
The ADC(2)-X method exhibits much larger \mae ($\sim$ 0.5 eV) due to its well-known unbalanced description of electron correlation effects.\cite{Dreuw:2014p82}

All methods show much higher errors in excitation energies for SSM. 
This is particularly noticeable for ADC(2), which \mae and \std grow $\sim$ 4-fold for both the UHF and ROHF references.
The ADC(3) method shows a more modest increase in \mae from WSM to SSM by a factor of $\sim$ 3.5 or $\sim$ 2.5 for when using the UHF or ROHF orbitals, respectively.
Overall, the \mae in excitation energies of SSM reduce in the order ADC(2)/UHF $>$ ADC(3)/UHF $>$ ADC(2)/ROHF $>$ ADC(3)/ROHF, which may be expected due to the higher-order description of electron correlation effects in ADC(3) and lower excited-state \SC in the calculations with  ROHF reference. 

A more detailed picture showing the role of spin contamination in ADC calculations of electronic excitations can be obtained by analyzing the data in \cref{fig:ee_error_vs_sc}, which plots the absolute errors in excitation energies ($\Delta \Omega_m$) against \SC for the excited states of SSM.
A linear regression analysis of this data reveals that, when using the UHF reference,  the ADC methods exhibit weak correlations between $\Delta \Omega_m$ and \SC with $0.11 \le R^2 \le 0.17$. 
The strongest correlation $R^2 = 0.17$ is observed for ADC(3)/UHF, suggesting that at this level of theory \SC plays a significant role in determining the errors in excitation energies, together with the approximations in description of electron correlation.
The ADC(2) method shows a weaker correlation with $R^2  = 0.11$, which indicates the increased role of approximate electron correlation treatment in determining $\Delta \Omega_m$. 

For ADC(2)/UHF, the $\Delta \Omega_m$ -- \SC correlation is negative, leading to smaller errors for the excited states with larger \SC. 
Similar trend has been observed in the ADC(2)/UHF calculations of charged excitation energies where the negative correlation has been attributed to the partial error cancellation originating from spin contamination and approximations in electron correlation treatment.\cite{stahl2022quantifying}
As discussed in \cref{sec:results:osc_strengths}, this reduction in excitation energy errors at the ADC(2)/UHF level of theory comes at a price of significantly less accurate oscillator strengths.
The ADC methods with ROHF reference show little to no correlation between $\Delta \Omega_m$ and \SC  in the linear regression analysis ($R^2 \approx 0.00, 0.04, 0.01$ for ADC(2), ADC(2)-X, ADC(3), respectively), suggesting that the errors of ADC methods with this reference primarily originate from approximations in the description of electron correlation. 
Overall, our analysis of data in \cref{fig:ee_error_vs_sc} reveals that \SC plays secondary but important role in determining the errors in excitation energies of SSM in the ADC calculations with UHF reference.

A particularly challenging test for the single-reference ADC methods is simulating electronic transitions between the spatially degenerate ground and excited states, such as the $X^2\Pi \rightarrow {1}^2\Pi$ excitations in CNO, NCO, and CON, as well as the $X^2\Pi \rightarrow {1}^2\Delta$ transition in CH. 
In these cases, using a single-determinant reference wavefunction breaks the spatial degeneracy of reference $\pi$-orbitals leading to different excitation energies when an electron is excited from (or into) either $\pi_x$ or $\pi_y$.
As reported in \cref{tab:ADC_EE_TZ_energies_strong}, the ADC calculations show strong spatial symmetry breaking lifting the degeneracy of ${1}^2\Pi$ or ${1}^2\Delta$ excited states in CNO, NCO, CON, and CH by up to $\sim$ 1 eV. 
This severe violation of spatial symmetry is accompanied by strong \SC reported in \cref{tab:ADC_EE_s_2_tz_strong} for each non-degenerate microstate.
While using the ROHF reference helps in lowering \SC compared to the calculations with UHF reference, it does not systematically reduce the energy difference between the spatially broken states.
Similarly, increasing the level of theory from ADC(2) to ADC(3) does not lead to consistent improvements in describing spatial degeneracy, although the excitation energies from ADC(3) tend to be in a better agreement with the reference data from full configuration interaction.
These results caution against using the single-reference ADC methods for simulating electronic excitations between the spatially degenerate ground and excited states.
The correct description of spin and spatial symmetry for these electronic transitions can be achieved using the multireference ADC methods.\cite{Sokolov:2018p204113,Mazin:2021p6152}

\subsection{Spin contamination and errors in oscillator strengths}
\label{sec:results:osc_strengths}

\begin{figure*}[t!]
	\centering
	\captionsetup{justification=raggedright,singlelinecheck=false,font=footnotesize}
	\includegraphics[scale=0.46,trim=0.0cm 0.0cm 0.0cm 0.0cm,clip]{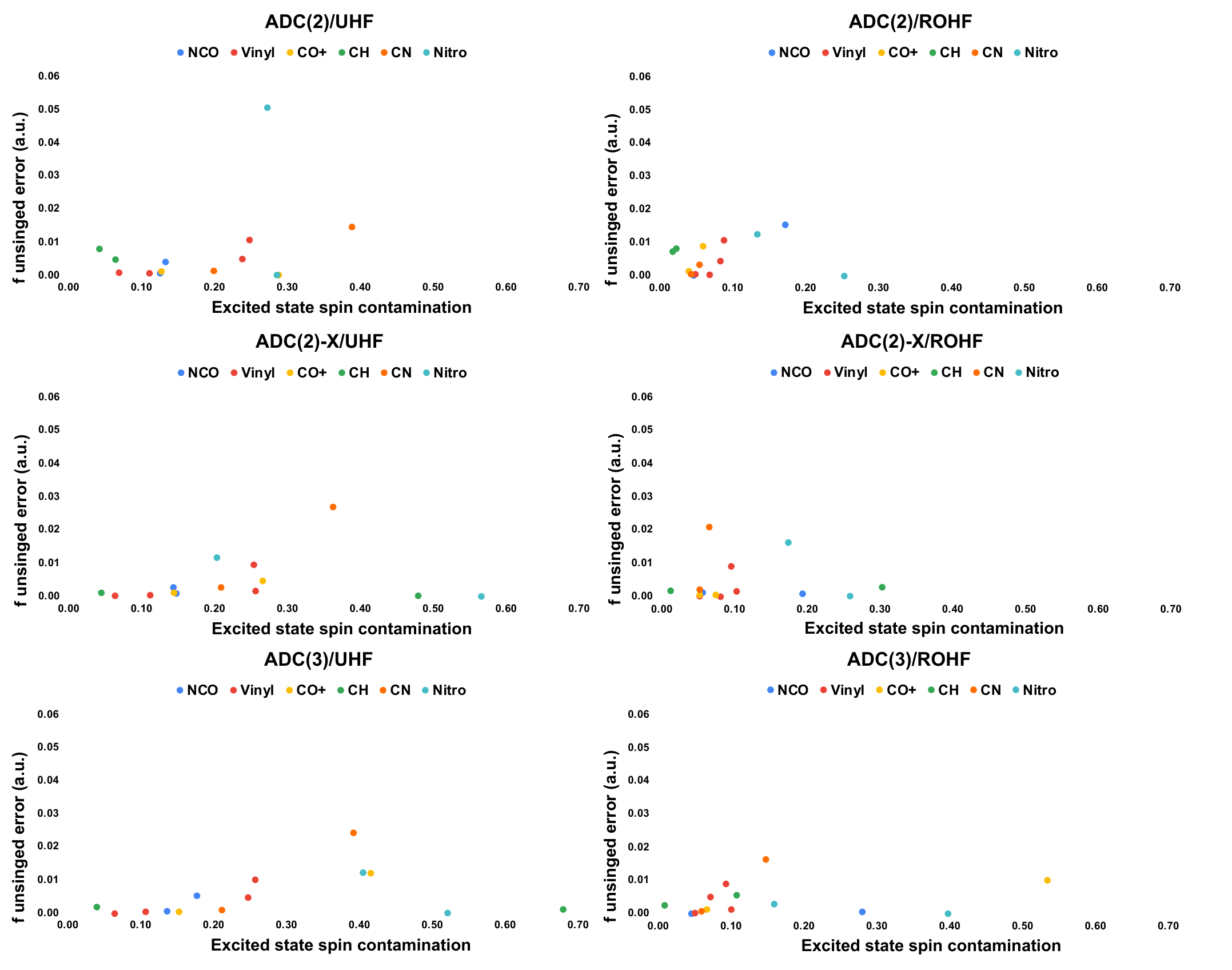}
	\caption{
		Unsigned errors in the oscillator strengths of strongly spin-contaminated molecules plotted against the excited-state spin contamination computed using the ADC methods with the UHF or ROHF reference wavefunctions.
		Calculations were performed using the aug-cc-pVDZ basis.
		See \cref{sec:Imp/comp_details} and Supplementary Materials for details.
 	}
	\label{fig:osc_strength_error_vs_sc}
\end{figure*}

In addition to excitation energies, reliable simulations of UV/Vis spectra require accurate calculations of transition dipole moments and oscillator strengths.
Recently, several benchmark studies of oscillator strengths have been reported for a variety of excited-state electronic structure methods\cite{jacquemin2016assessment,chrayteh2020mountaineering,sarkar2021benchmarking,damour2022ground} 
including ADC.\cite{harbach2014third,li2014comparison,wormit2014investigating,schutz2015oscillator,maier2023consistent}
All of these studies, however, focused on excited electronic states of closed-shell molecules, which can be simulated with little to no spin contamination by starting with a restricted reference wavefunction. 

Here, we assess the accuracy of ADC oscillator strengths ($f_m$) for open-shell molecules with strong spin contamination in excited states.
To quantify the errors in $f_m$ ($\Delta f_m$), we computed accurate oscillator strengths using the ROHF-based linear-response coupled cluster theory with up to triple excitations (LR-CCSDT/ROHF) and the aug-cc-pVDZ basis set. 
Due to the challenges associated with performing LR-CCSDT/ROHF calculations, only 6 out of 9 SSM from \cref{sec:results:energies} were included in this study.
The ADC and LR-CCSDT/ROHF oscillator strengths are reported in the Supplementary Material.
For the spatially broken $X^2\Pi \rightarrow {1}^2\Pi$ excitations in NCO, only one of the two non-degenerate transitions has non-zero oscillator strength, which was used to calculate $\Delta f_m$ relative to LR-CCSDT/ROHF.
For the $X^2\Pi \rightarrow {1}^2\Delta$ excitation in CH, $\Delta f_m$ was computed for each of the two spatially broken microstates with non-zero oscillator strength separately. 

\cref{fig:osc_strength_error_vs_sc} plots $\Delta f_m$ against \SC for each electronic transition.
The UHF-based ADC methods exhibit positive correlations between $\Delta f_m$  and \SC in the linear regression analysis, suggesting that the excited states with larger \SC are more likely to show large errors in $\Delta f_m$ than those with smaller \SC.
The strongest correlations are observed for ADC(2) ($R^2 = 0.12$) and ADC(3) ($R^2 = 0.10$), while for ADC(2)-X this trend is less pronounced ($R^2 = 0.04$).
Using the ROHF reference lowers the average and maximum error in $f_m$ for all ADC methods.
In addition, the ROHF-based ADC methods show significantly weaker $\Delta f_m$  -- \SC correlations with $R^2$ ranging from 0.00 to 0.06.
Overall, these results highlight that the correct description of electronic spin is important for accurate calculations of oscillator strengths and provide evidence that the transition properties calculated using the ADC methods can be susceptible to spin contamination, particularly at low perturbation orders.

\subsection{Case study: spin contamination in the UV/Vis spectrum of phenyl radical}
\label{sec:results:phenyl}

Having analyzed the effect of spin contamination on the ADC excitation energies and oscillator strengths, we now take a closer look at how this unphysical spin symmetry breaking manifests itself in the ADC simulations of UV/Vis spectra. 
For this study, we focus on phenyl radical ($\ce{C6H5}$), which plays an important role in organic synthesis, environmental chemistry, combustion reactions, and chemical processes in interstellar objects.\cite{o1994involvement,imoto2008impaired,sekine2008role,frenklach1989formation,gu2009reaction,jones2011formation,kaiser2015reaction,parker2017formation}  
Despite being the smallest aromatic radical,\cite{martinez2015accurate} $\ce{C6H5}$ has a nontrivial electronic structure in its $\tilde{X}\,^{2}A_{1}$ ground state with a heavily spin-contaminated UHF wavefunction ($\SC =$ 0.40 a.u.\@ at the MP2/aug-cc-pVTZ optimized geometry), which presents challenges for simulating its UV/Vis spectrum using single-reference electronic structure methods.\cite{blanquart2015effects}

\begin{figure*}[t!]
	\centering
	\captionsetup{justification=raggedright, singlelinecheck=false, font=footnotesize}
	\includegraphics[scale=0.75, trim=0.0cm 0.0cm 0.0cm 0.0cm, clip]{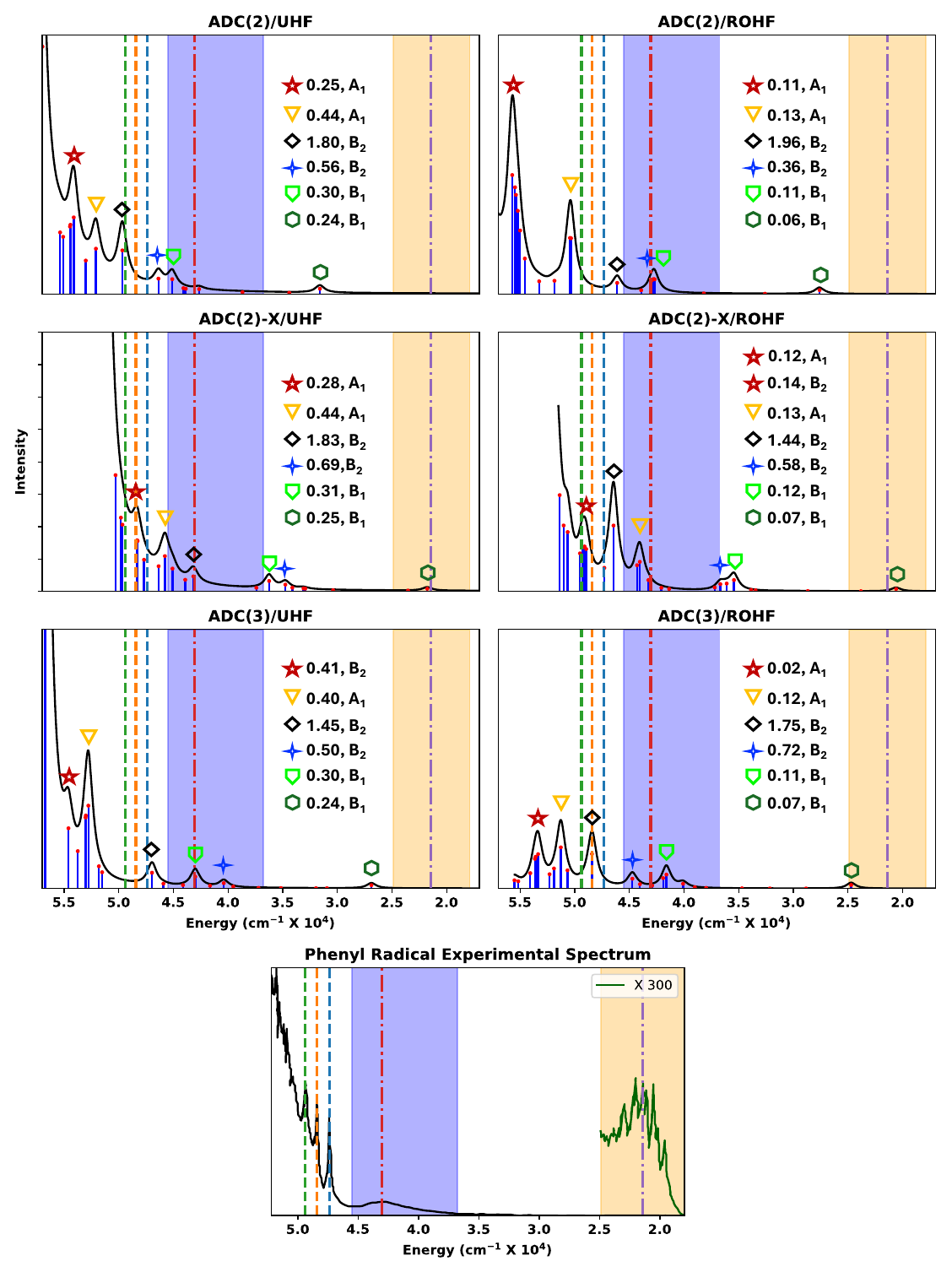}%
	\caption{The UV/Vis spectra of phenyl radical measured experimentally\cite{radziszewski1999electronic} and simulated using the ADC methods.
	Prominent features in the experimental spectrum are highlighted with shading or dashed vertical lines. 
	For the peaks labeled in simulated spectra, the spin contamination and spatial symmetry of corresponding excited states are provided in the legend.
	See \cref{sec:Imp/comp_details} and Supplementary Materials for  computational details and complete data.
	}
	\label{fig:phenyl_results}
\end{figure*}

Several experimental studies of phenyl radical UV/Vis absorption have been reported.\cite{porter1965electronic,cercek1970phenyl,miller1980argon,ikeda1985observation,hatton1990photochemistry,engert1996uv,tonokura2002cavity,song2012ultraviolet} 
The most complete UV/Vis spectrum of $\ce{C6H5}$ in the 52632 -- 4000 cm$^{-1}$ range was measured in a noble gas matrix by Radziszewski\cite{radziszewski1999electronic} and is shown in \cref{fig:phenyl_results}.
In this spectrum, three regions can be identified: 1) a very weak broad feature between 25000 and 17500 cm$^{-1}$ centered at $\sim$ 21500 \cm (shaded orange), 2) a broad (45000 -- 40000 cm$^{-1}$) band with medium intensity and a maximum at $\sim$ 43000 \cm (shaded blue), and 3) a higher-energy region ($>$ 46000 cm$^{-1}$) with three intense peaks at 47365, 48430, and 49838 \cm (shown as dashed lines).

Extensive theoretical studies of $\ce{C6H5}$ UV/Vis spectra and excited states have been also  performed.\cite{pacansky1983photolysis,johnson1984ab,krauss1994excited,engert1996uv,kim2002ab,starcke2009unrestricted,biczysko2009first} 
While all reported calculations were in agreement with the experiments about the nature of electronic transition in region 1, some of them provided different interpretations of spectral features in region 2 and 3.\cite{kim2002ab,starcke2009unrestricted} 
In addition, the simulated UV/Vis spectra revealed strong dependence on the level of theory used in calculations.

Here, we investigate the phenyl radical UV/Vis spectrum using the ADC methods and the large aug-cc-pVTZ basis set, paying particular attention to the spin contamination in its excited states. 
To organize our discussion, we focus on each of three regions in \cref{fig:phenyl_results} one at a time, starting with the lowest-energy electronic transition.

\subsubsection{Region 1 (25000 -- 17500 \cm)}

This region of experimental UV/Vis spectrum (\cref{fig:phenyl_results}) consists of a broad band with low intensity centered at $\sim$ 21500 \cm.\cite{radziszewski1999electronic}
Previous experimental and theoretical studies assigned this spectral feature as excitation from the ground $\tilde{X}\,^{2}A_{1}$ state to an excited state of ${}^{2}B_{1}$ symmetry.\cite{porter1965electronic,pacansky1983photolysis,johnson1984ab,krauss1994excited,tonokura2002cavity,kim2002ab,starcke2009unrestricted,biczysko2009first}
In agreement with this assignment, the ADC-simulated UV/Vis spectra (\cref{fig:phenyl_results}) show a low-energy transition to the ${}^{2}B_{1}$ state with a small oscillator strength. 
When combined with the UHF reference, the ADC(2) method strongly overestimates the excitation energy for this state, predicting it to be $\sim$ 6500 \cm outside the experimental range.
The ADC(2)-X/UHF and ADC(3)/UHF methods show better agreement with the experiment.
All UHF-based ADC calculations exhibit significant spin contamination in the ${}^{2}B_{1}$ excited state with \SC $\sim$ 0.25 a.u.

Using the ROHF reference reduces the ${}^{2}B_{1}$ spin contamination by a factor of three and lowers its excitation energy by $\sim$ 4000 \cm for ADC(2), $\sim$ 1200 \cm for ADC(2)-X, and $\sim$ 2000 \cm for ADC(3), significantly improving their agreement with the experimental results. 
Reducing spin contamination lowers the oscillator strength computed using ADC(2), but has a minor effect on transition intensities obtained from ADC(2)-X and ADC(3).
Although the ADC(2)-X/ROHF excitation energy appears to be in a better agreement with the experiment than ADC(3)/ROHF, we note that the simulated spectra reflect vertical excitation energies and that including excited-state structural relaxation and vibrational effects is expected to improve the performance of ADC(3)/ROHF relative to ADC(2)-X/ROHF. 

\subsubsection{Region 2 (45000 -- 40000 \cm)}

In this region, the experimental UV/Vis spectrum (\cref{fig:phenyl_results}) exhibits a broad signal with medium intensity and a maximum at $\sim$ 43000 \cm.\cite{radziszewski1999electronic}
Although this feature was experimentally assigned as an excitation to the ${}^{2}A_{1}$ state, theoretical studies suggested that the absorption in this region is dominated by the electronic excitations to either ${}^{2}B_{1}$  or ${}^{2}B_{2}$, with some disagreement between different levels of theory.\cite{kim2002ab,starcke2009unrestricted} 
Indeed, the ADC-simulated UV/Vis spectra reported in \cref{fig:phenyl_results} exhibit several peaks near the 45000 -- 40000 \cm region with the most intense transitions identified as excitations to the ${}^{2}B_{1}$  or ${}^{2}B_{2}$ states.
The relative energies, intensities, and order of these states, however, strongly depend on the level of theory and reference wavefunction.

The ADC(2)/UHF method predicts very similar intensities for the ${}^{2}B_{1}$  and ${}^{2}B_{2}$ transitions and overestimates their energies relative to the experimental results. 
Increasing the level of electron correlation treatment to ADC(3)/UHF lowers the ${}^{2}B_{1}$  and ${}^{2}B_{2}$ excitation energies, changes their order in the energy spectrum, and decreases the intensity of ${}^{2}B_{2}$ relative to ${}^{2}B_{1}$.
Both ${}^{2}B_{1}$ and ${}^{2}B_{2}$  exhibit strong spin contamination with \SC $\sim$ 0.30 and 0.55 a.u., respectively.

Combining ADC(2) with the ROHF reference decreases \SC in both ${}^{2}B_{1}$ and ${}^{2}B_{2}$, lowers their excitation energy by up to $\sim$ 3000 \cm, and improves the agreement of simulations with experiment.
For ADC(3), reducing spin symmetry breaking in the reference wavefunction changes the peak structure in the 45000 -- 40000 \cm region where a new ${}^{2}B_{2}$ peak appears at higher energy ($\sim$ 45000 \cm), with intensity somewhat lower than that of ${}^{2}B_{1}$ and large \SC = 0.72 a.u.
Among all ADC methods employed in this work, ADC(3)/ROHF shows the best agreement with experiment in this region, despite exhibiting strong spin contamination in its results.  

\subsubsection{Region 3 ($>$ 46000 \cm)}
Finally, we consider the highest energy region ($>$ 46000 \cm) where the experimental UV/Vis spectrum shows three peaks with high intensity at $\sim$ 47360, 48430, and 49840 \cm, labeled with dashed lines in \cref{fig:phenyl_results}.\cite{radziszewski1999electronic}
The lowest-energy peak in this region (47360 \cm) has been experimentally assigned as a transition to the ${}^{2}B_{2}$ excited state,\cite{radziszewski1999electronic} but the theoretical studies using MRCI and ADC(2)-X/UHF with smaller basis sets than the one employed in this work labeled this excitation as ${}^{2}B_{1}$.\cite{kim2002ab,starcke2009unrestricted} 
No reliable assignments have been made for the two higher-energy features at 48430 and 49840 \cm.

The ADC-simulated UV/Vis spectra consistently show three intense peaks with transition energies $\gtrsim$ 45000 \cm, but the relative spacing, order, and intensities of the corresponding excited states vary significantly depending on the level of theory and reference wavefunction employed.
With the exception of ADC(2)/ROHF, all ADC calculations assign the lowest-energy peak in this region as a transition to the ${}^{2}B_{2}$ state, which agrees with the experimental interpretation\cite{radziszewski1999electronic} but contradicts prior theoretical assignments.\cite{kim2002ab,starcke2009unrestricted}  
Same methods also assign the second-energy feature as excitation to the ${}^{2}A_{1}$ state.
The assignment of the highest-energy peak is complicated due to the high density of states, but most methods reveal strong absorption from the higher-lying  ${}^{2}A_{1}$ with a significant contribution from ${}^{2}B_{2}$.

When using the UHF reference, all excited states in this region show strong spin contamination.
The highest degree of spin symmetry breaking is observed for the lowest-energy ${}^{2}B_{2}$ state where \SC ranges from 1.45 to 1.83 a.u.\@ depending on the level of theory used.
Starting the calculations with the ROHF reference has a profound effect on the simulated UV/Vis spectra, changing the order of peaks, their spacing, and relative intensities. 
Although reducing spin contamination in the reference wavefunction helps to lower \SC for the ${}^{2}A_{1}$ excited states, the lowest-energy ${}^{2}B_{2}$ state remains highly spin-contaminated.
Neither method produces UV/Vis spectra in satisfactory agreement with the experiment, although increasing the basis set size and incorporating effects of excited-state structural relaxation may improve the results. 
Nevertheless, the residual spin contamination in the calculated UV/Vis spectra is concerning and warrants the development of spin-adapted ADC methods for open-shell systems.

\section{Conclusions}
\label{sec:conclusions}

In this work, we performed a systematic study of spin contamination (SC) in the calculations of excited electronic states and UV/Vis spectra of open-shell molecules using algebraic diagrammatic construction theory (ADC).
As a perturbation theory, ADC offers a hierarchy of excited-state methods that can deliver accurate results more efficiently than the non-perturbative approaches such as coupled cluster theory. 
These computational savings come at a cost of higher sensitivity to the quality of reference wavefunction, which can be especially problematic for open-shell systems.

Our work reports several important findings about the role of SC in the ADC calculations of excitation energies, oscillator strengths, and UV/Vis spectra.
First, we demonstrate that the accuracy of ADC methods strongly depends on the level of SC in the UHF reference wavefunction.
In particular, for open-shell molecules with the UHF SC $\le$ 0.05 a.u., the second-order (ADC(2)) and third-order (ADC(3)) ADC methods exhibit high accuracy with the errors of $\sim$ 0.1 eV in excitation energy, on average.
However, when applied to strongly spin-contaminated molecules with the UHF SC $>$ 0.05 a.u., their performance significantly deteriorates. 
One of the factors contributing to this poorer performance is the SC in excited states, which increases significantly as the reference SC grows.
Performing the ADC calculations with the ROHF reference wavefunctions allows to significantly reduce the level of SC in the excited states, but does not eliminate it entirely. 
Increasing the level of theory from ADC(2) to ADC(3) does not help reducing the spin symmetry breaking in excited states, and may lead to increasing SC instead.

Our second finding is that SC can significantly affect the accuracy of ADC oscillator strengths and calculated UV/Vis spectra.
In particular, we demonstrated that the molecules with strong SC in the UHF reference tend to show larger errors in ADC oscillator strengths compared to the molecules with lower reference and excited-state SC.
To investigate this further, we performed a study of ADC-simulated UV/Vis spectra for the phenyl radical that exhibits large SC in the UHF reference wavefunction.
Using the ROHF reference improves the agreement of simulated spectra with the experiment at the ADC(3) level of theory, but the computed excited states still show significant levels of SC. 
We note that, in comparison to ADC with the UHF reference, the ROHF-based ADC calculations of UV/Vis spectra provide a higher-level description of coupling between the ground state and singly-excited electronic configurations, which may be responsible for some of the observed improvements in the simulated UV/Vis spectra in addition to lowering SC.\cite{Sulzner:2024p2462}

Overall, the results of our work suggest that the unrestricted implementations of ADC methods should be used with caution when applied to molecules with the UHF SC  $>$ 0.05 a.u.\@ 
If such calculations need to be performed, we recommend to employ the ROHF reference wavefunction, which allows to reduce SC but does not eliminate it entirely.
To address the SC problem, spin-adapted implementations of single-reference ADC methods that preserve spin symmetry for open-shell systems need to be developed.

\section{Supplementary Material}
See the supplementary material for the ADC results computed using the aug-cc-pVDZ basis set, the ADC and  LR-CCSDT oscillator strengths, the ADC spectral data for the phenyl radical, and equations for the ADC(2) one- and two-particle reduced density matrices.

%%%%%%%%%%%%%%%%%%%%%%%%%%%%
% Acknowledgements
%%%%%%%%%%%%%%%%%%%%%%%%%%%%
\section{Acknowledgements}
Acknowledgment is made to the donors of the American Chemical Society Petroleum Research Fund for supporting this research (PRF 65903-ND6).
Computations were performed at the Ohio Supercomputer Center under projects PAS1963.\cite{OhioSupercomputerCenter1987} 

\section{Data availability}
The data that supports the findings of this study are available within the article and its supplementary material. 
Additional data can be made available upon reasonable request.

%\bibliography{refs_terry,terry_ref,papers_alex}

\end{document}